\documentclass[pre,superscriptaddress,twocolumn]{revtex4}
\usepackage{graphicx}
\usepackage{lipsum}
\usepackage{float}
\usepackage{amsmath}
\usepackage{amssymb}
\usepackage{mathtools}

\begin{document}

\title{Statistical analysis of edges and bredges \\
in configuration model networks}

\author{Haggai Bonneau}
\affiliation{Racah Institute of Physics, The Hebrew University, Jerusalem 9190401, Israel}

\author{Ofer Biham}
\affiliation{Racah Institute of Physics, The Hebrew University, Jerusalem 9190401, Israel}

\author{Reimer K\"uhn}
\affiliation{Department of Mathematics, King's College London, Strand, London WC2R 2LS, UK}

\author{Eytan Katzav} 
\affiliation{Racah Institute of Physics, The Hebrew University, Jerusalem 9190401, Israel}

\begin{abstract}
A bredge (bridge-edge) in a network is an edge whose deletion would
split the network component on which it resides into two separate
components.
Bredges are vulnerable links that play an important role in network collapse
processes, which may result from node or link failures, attacks or epidemics.
Therefore, the abundance and properties of bredges affect the resilience
of the network to these collapse scenarios.
We present analytical results for the statistical properties of 
bredges in configuration model networks.
Using a generating function approach
based on the cavity method,
we calculate the probability
$\widehat P(e \in {\rm B})$
that a random edge $e$ in a configuration model network 
with degree distribution $P(k)$
is a bredge (B).
We also calculate the joint degree distribution  
$\widehat P(k,k' | {\rm B})$
of the end-nodes $i$ and $i'$ of a random bredge.
We examine the distinct properties of bredges on the giant component (GC)
and on the finite tree components (FC) of the network.
On the finite components all the edges are bredges 
and there are no degree-degree correlations.
We calculate the probability
$\widehat P(e \in {\rm B}|{\rm GC})$ 
that a random edge on the giant component is a bredge. 
We also calculate the joint degree distribution
$\widehat P(k,k'|{\rm B},{\rm GC})$
of the end-nodes of bredges
and the joint degree distribution 
$\widehat P(k,k'|{\rm NB},{\rm GC})$
of the end-nodes of non-bredge (NB) edges
on the giant component.
Surprisingly, it is found that the degrees $k$ and $k'$ of the end-nodes of bredges are correlated,
while the degrees of the end-nodes of non-bredge edges are uncorrelated.
We thus conclude that all the degree-degree correlations on the giant component
are concentrated on the bredges.
We calculate the covariance
$\Gamma({\rm B},{\rm GC})$ 
of the joint degree distribution of end-nodes of bredges
and show it is negative, namely bredges tend
to connect high degree nodes to low degree nodes.
We apply this analysis to ensembles of configuration model networks
with degree distributions that follow 
a Poisson distribution (Erd{\H o}s-R\'enyi networks), 
an exponential distribution 
and a power-law distribution 
(scale-free networks).
The implications of these results are discussed in the context of common
attack scenarios and network dismantling processes.
\end{abstract}

\maketitle

\section{Introduction}

Network models provide a useful conceptual framework 
for the study of a large variety of systems and processes
in science, technology and society
\cite{Havlin2010,Newman2010,Estrada2011,Barrat2012,Latora2017}.
These models consist of nodes and edges, where the nodes
represent physical objects, while the edges represent the
interactions between them.
Unlike regular lattices in which all the nodes have the same coordination
number, network models are characterized by a degree distribution 
$P(k)$.
The backbone of a network often consists of high degree
nodes or hubs, which connect the different branches and
maintain the integrity of the network.
In some applications,
such as communication networks,
it is crucial that the network will consist of a single connected component.
However, mathematical models also produce networks
that combine a giant component and isolated finite components,
as well as fragmented networks that consist only of isolated finite components
\cite{Bollobas2001}.

Networks are often exposed to the loss of nodes and edges, which may
severely affect their functionality. 
Such losses may occur due to inadvertent node or edge failures,
propagation of epidemics or deliberate attacks.
Starting from a single connected component,
as nodes or edges are deleted they may lead to the separation of network
fragments from the giant component. 
As a result, the size of the giant component decreases until it 
completely disintegrates.
The ultimate failure, when the network fragments into isolated finite
components was studied extensively using percolation theory
\cite{Albert2000,Cohen2000,Cohen2001,Schneider2011}.

A major factor in the sensitivity of networks to node or edge deletion processes is
the fact that the deletion of a single node or a single edge may separate a whole fragment
from the giant component. 
This fragmentation process greatly accelerates
the disintegration of the network.
Using iterative search algorithms
one can identify the nodes whose deletion would break
the component on which they reside into two or more components
\cite{Hopcroft1973,Gibbons1985,Chaudhuri1998}
Such nodes, called articulation points (APs), 
were recently studied in the context of network resilience and optimized attack strategies
\cite{Tian2017}.
Using similar methods one can also identify 
the edges whose deletion would
break the component on which they reside into two separate components
\cite{Tarjan1974,Italiano2012}.
Such edges are called bridge-edges or cut-edges
\cite{Bollobas1998}.
Here we use the term {\it bredges},
which provides a shorthand for bridge-edges,
and avoids a potential confusion with many other technical terms involving
the word `bridge'.
Moreover, the word 'bredge' was used in ancient English as a synonym
to the word `bridge'
\cite{Bredge1685}.
In fact, an edge that does not participate in any cycle is a bredge (B).
Thus, in network components that exhibit a tree structure,
such as the finite tree components of configuration model networks,
all the edges are bredges.


\begin{figure}
\includegraphics[width=5cm]{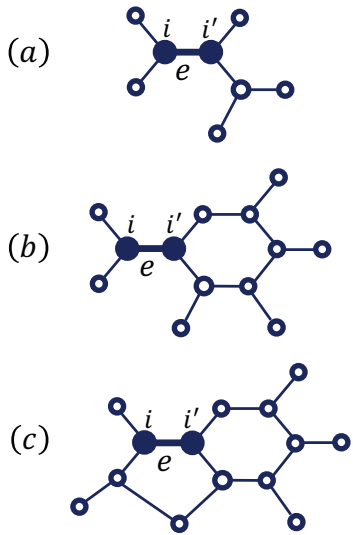} 
\caption{
Schematic illustration of bredges and their surrounding network components:
(a) A bredge $e$ (marked by a thick line) in a finite tree component.
Deletion of the bredge $e$ would split the tree component into two separate tree components.
The end-node $i$ will reside on one of the tree components and the end-node $i'$ will reside on the
other tree component;
(b) A bredge $e$ (thick line) where one of its end nodes, $i'$, resides on a cycle.
Deletion of the bredge would split the network into two separate components;
(c) Here the edge $e$, marked by a thick line, is not a bredge because the end nodes of this
edge are connected by another path. As a result, upon deletion of
the marked edge its two end nodes remain on the same network component.
}
\label{fig:1}
\end{figure}

In Fig. \ref{fig:1}(a) we present a
schematic illustration of a bredge $e$ (marked by a thick line) 
in a tree network and its end-nodes $i$ and $i'$ (full circles).
Deletion of the bredge would split the network into two separate tree components.
In Fig. \ref{fig:1}(b) we show a bredge $e$ (thick line), 
where one of its end-nodes, $i'$, resides on a cycle.
Deletion of the bredge would split the network into two separate components.
The component that includes the end-node $i'$ represents the giant component of the 
reduced network from which $e$ is removed, while the component that includes
the end-node $i$ represents the finite tree component that is detached 
from the giant component upon deletion of $e$.
The edge $e$ marked by a thick line in Fig. \ref{fig:1}(c) 
is not a bredge because its end-nodes are
connected by a path that does not go through $e$. 
As a result, upon deletion of
$e$ its end-nodes $i$ and $i'$ remain on the same network component.
Since the paths connecting the end-nodes of an edge $e$ may be long,
the determination of whether $e$ is a bredge or not cannot be done locally and 
requires access to the large-scale structure of the whole network
\cite{Hopcroft1973,Gibbons1985}.

In practice, the functionality of most networks relies on the integrity of their giant components.
Therefore, it is particularly important to study the properties of bredges and APs that reside on the 
giant component. These bredges and APs are vulnerable spots in the structure of a network,
because the deletion of a single bredge may detach an entire tree branch 
from the giant component while the deletion of a single AP may detach 
one or several tree branches.
This vulnerability is exploited in network attack strategies, which 
generate new bredges and AP via decycling processes 
and then attack them to dismantle the network
\cite{Tian2017,Braunstein2016,Zdeborova2016,Wandelt2018,Ren2019}.
While bredges and AP make the network vulnerable to attacks, they are advantageous
in fighting epidemics. 
In particular, maintaining isolation between nodes
connected by bredges prevents the spreading of epidemics between the network
components connected by these bredges.
Similarly, in communication networks the party in possession of an AP or a bredge
may control, screen, block or alter the communication between the network components
connected by the AP or the bredge.

There is an intricate connection between bredges and APs.
On the one hand, each one of the end-nodes $i$ and $i'$ of a bredge $e$ is either an AP 
(if its degree satisfies $k \ge 2$) or a leaf node (if its degree is $k=1$).
On the other hand, if a node $i$ of degree $k \ge 2$ is an AP, then
at least one of its $k$ edges must be a bredge.
Moreover, in the case of 
a node $i$ of degree 
$k=2$, both edges of $i$ are bredges.
The statistical properties of APs in 
configuration model networks 
were studied in a recent paper
\cite{Tishby2018a}.
The probability
$P(i \in {\rm AP})$
that a random node $i$ in a configuration model network 
with degree distribution $P(k)$
is an AP was calculated. 
Moreover, closed form expressions were obtained for
the conditional probability 
$P(i \in {\rm AP} | k)$
that a random node of a given degree $k$ is an AP and for the
conditional degree distribution 
$P(k | {\rm AP})$.
An important property of an AP is the articulation rank $r$, which is
the number of components 
that are added to the network upon deletion of the AP.
For each node in the network the articulation rank
satisfies $0 \le r \le k$, where $k$ is the degree of the node.
The articulation rank of a node which is not an AP is $r=0$, while the
articulation ranks of APs satisfy $r \ge 1$.
In fact, the articulation rank of an AP is the number of bredges
connected to it.
The distribution 
$P(r)$ 
of articulation ranks 
was calculated in Ref. 
\cite{Tishby2018a}.

In this paper
we present analytical results for the statistical properties of 
bredges in configuration model networks.
In order to quantify the abundance of bredges,
we calculate the probability
$\widehat P(e \in {\rm B})$,
that a random edge $e$ in a configuration model network 
with degree distribution $P(k)$ 
is a bredge.
To characterize the statistical properties of bredges, 
we derive a closed form expression for
the joint degree distribution  
$\widehat P(k,k' | {\rm B})$
of the end-nodes $i$ and $i'$ of a random bredge.
We also examine the distinct properties of bredges on the giant component (GC)
and on the finite tree components (FC) of the network.
On the finite components all the edges are bredges,
namely $\widehat P(e \in {\rm B}|{\rm FC})=1$.
We calculate the probability 
$\widehat P(e \in {\rm B}|{\rm GC})$
that a random edge that resides on
the giant component is a bredge
and the joint degree distribution
$\widehat P(k,k'|{\rm B},{\rm GC})$
between the end-nodes of bredges on the giant component.
It is found that the degrees $k$ and $k'$ of the end-nodes of a bredge 
that resides on the giant component are correlated.
This is in contrast to the end-nodes of random edges in the network
and to the end-nodes of non-bredge (NB) edges on the giant component, 
which exhibit no degree-degree correlations.
We thus conclude that all the degree-degree correlations on the giant component
are concentrated on the bredges.
We calculate the covariance
$\Gamma({\rm B},{\rm GC})$ 
and show that it is negative, which means that bredges 
on the giant component tend
to connect high degree nodes to low degree nodes.
We apply these results to ensembles of configuration model networks
with degree distributions that follow 
a Poisson distribution (Erd{\H o}s-R\'enyi networks), 
an exponential distribution 
and a power-law distribution 
(scale-free networks).

The paper is organized as follows.
In Sec. II we describe the configuration model network and its construction.
In Sec. III we present the generating functions of the degree distribution.
In Sec. IV we present a statistical analysis of nodes on the giant component and on the 
finite components.
In Sec. V we present a statistical analysis of edges on the giant and finite components.
In Sec. VI we present a detailed statistical analysis of bredges.
In Sec. VII we apply these results to configuration model networks with  
a Poisson degree distribution (ER networks), exponential degree distribution
and power-law degree distribution (scale-free networks).
The results are discussed in Sec. VIII and summarized in Sec. IX.

\section{The configuration model}

The configuration model is an ensemble 
of uncorrelated random networks
whose degree sequences are drawn from
a given degree distribution $P(k)$.
The first moment (mean degree) and the second moment 
of $P(k)$ are denoted by
%
$\langle K^n \rangle$,
%
where $n=1$ and $2$, respectively,
while the variance is given by
${\mathbb V}[K] = \langle K^2 \rangle - \langle K \rangle^2$.
The support of the degree distribution
of random networks is often bounded from below
by $k_{\rm min} \ge 1$ such that $P(k)=0$
for $0 \le k \le k_{\rm min}-1$,
with non-zero values of
$P(k)$ only for
$k \ge k_{\rm min}$.
For example, the commonly used choice of $k_{\rm min}=1$
eliminates the possibility of isolated nodes in the network. 
Choosing $k_{\rm min}=2$ also eliminates the leaf nodes.
One may also control the upper bound by imposing
$k \le k_{\rm max}$.
This may be important in the case of finite networks
with heavy-tail degree distributions  
such as power-law distributions.
The configuration model network ensemble is a maximum entropy ensemble
under the condition that the degree distribution $P(k)$ is imposed
\cite{Molloy1995,Molloy1998,Newman2001}.
Here we focus on the case of undirected networks.

To generate a network instance drawn from an ensemble of
configuration model networks of $N$ nodes,
with a given degree distribution $P(k)$, one draws
the degrees of the $N$ nodes independently from 
$P(k)$.
This gives rise to a
degree sequence of the form
$k_1,k_2,\dots,k_N$.
For the discussion below it is convenient to list the degree
sequence in a decreasing order of the form
$k_1 \ge k_2 \ge \dots \ge k_N$.
It turns out that not every possible degree sequence is graphic,
namely admissible as a degree sequence of a network.
Therefore, before trying to construct a network with a given
degree sequence, one should first confirm
the graphicality of the degree sequence.
To be graphic, a degree sequence must satisfy two conditions.
The first condition is that the sum of the degrees is an even number,
namely
$\sum_{i} k_i = 2 L$,
where $L$ is an integer that represents
the number of edges in the network.
The second condition is expressed by the Erd{\H o}s-Gallai theorem,
which states that an ordered sequence of the form
$k_1 \ge k_2 \ge \dots \ge k_N$
that satisfies the first condition
is graphic if and only if the condition
\begin{equation}
\sum_{i=1}^n k_i \le n(n-1) + \sum_{i=n+1}^N \min (k_i,n)
\label{eq:EG}
\end{equation}

\noindent
holds for all values of $n$ in the range
$1 \le n \le N-1$
\cite{Erdos1960b,Choudum1986}.

A convenient way to construct a configuration model network 
is to prepare the $N$ nodes such that each node $i$ is 
connected to $k_i$ half edges or stubs
\cite{Newman2010}.
At each step of the construction, one connects a random pair of stubs that 
belong to two different nodes $i$ and $j$ 
that are not already connected,
forming an edge between them.
This procedure 
is repeated until all the stubs are exhausted.
The process may get stuck before completion in case that
all the remaining stubs belong to the
same node or to pairs of nodes that are already connected.
In such case one needs to perform some random reconnections
in order to complete the construction.

In the dense-network limit, configuration model networks consist of 
a single connected component, while in the dilute-network limit they 
consist of many finite tree components. 
At intermediate densities they exhibit a coexistence between a giant 
component, which is extensive in the network size, and 
many non-extensive finite tree components.
Some commonly studied configuration model networks can be 
described in terms of single parameter families of degree distributions.
A particularly convenient choice of the parameter is the mean degree
$c=\langle K \rangle$. 
In this case, the degree distribution can be expressed by
$P(k)=P_c(k)$, such that small values of $c$ correspond to
the dilute network limit while large values of $c$ correspond to the
dense network limit.
At some value $c_0$, referred to as the percolation threshold,
there is a percolation transition below which
the network consists of finite tree components and above which 
a giant component emerges.
The percolation transition is a second order phase transition, 
whose order parameter is the fraction $g$ of nodes that reside on
the giant component.
Below the transition, where $c < c_0$, the order parameter is $g=0$,
while for $c > c_0$ the function $g=g(c)$ gradually increases.

\section{The generating functions of the degree distribution}

Consider a configuration model network 
with a given degree distribution $P(k)$.
To obtain the probability $g$ that a random node
in the network belongs to the giant component,
one needs to first calculate the probability $\tilde g$,
that a node $i$ selected via a random edge $e$
belongs to the giant component of the reduced network, 
from which the edge $e$ is removed.
The probability $\tilde g$ is determined by
\cite{Havlin2010,Newman2010}
\begin{equation}
1 - {\tilde g} = G_1(1 - {\tilde g}),
\label{eq:tg}
\end{equation}

\noindent
where
\begin{equation}
G_1(x) = \sum_{k=1}^{\infty}   x^{k-1}   {\widetilde P}(k) 
\label{eq:G1}
\end{equation}

\noindent
is the generating function of the distribution
\begin{equation}
{\widetilde P}(k) = \frac{ k }{ \langle K \rangle } P(k), 
\label{eq:tPk}
\end{equation}

\noindent
which is the degree distribution of nodes that are sampled
via random edges.
The solution of Eq. (\ref{eq:tg}) is an attractive fixed point
(Sec. 13.8 in Ref. \cite{Newman2010}).
Using $\tilde g$, one can then obtain the probability $g$ from the equation
\begin{equation}
g = 1 - G_0(1 - {\tilde g}),
\label{eq:g}
\end{equation}

\noindent
where
\begin{equation}
G_0(x) = \sum_{k=0}^{\infty}   x^{k}   P(k) 
\label{eq:G0}
\end{equation}

\noindent
is the generating function of the degree distribution $P(k)$.
The two generating functions are related to each other by
%
$G_1(x) =    G'_0(x) /  G'_0(1) $,
%
where $G'_0(x)$ is the derivative of $G_0(x)$.

From the definitions of $G_0(x)$ and $G_1(x)$ in Eqs.
(\ref{eq:G0})
and 
(\ref{eq:G1}),
respectively,
we find that
$0 < G_0(x),G_1(x) < 1$
for $0 < x < 1$
and
$G_0(1)=G_1(1)=1$
This 
means that $x=1$ is a fixed point for both generating functions.
Therefore, $g=\tilde g=0$ is a solution of Eqs. (\ref{eq:tg}) and (\ref{eq:g}).
This solution corresponds to the case of subcritical networks,
in which there is no giant component.
In some networks there are no isolated nodes (of degree $k=0$)
and no leaf nodes (of degree $k=1$). 
In such networks $P(0)=0$ and $P(1)=0$, while
$P(k) > 0$ only for $k \ge 2$. 
The generating functions associated with these networks satisfy
$G_0(0) = 0$ 
and
$G_1(0)=0$.
This implies that in such networks both $x=0$ and $x=1$ are fixed points
of both $G_0(x)$ and $G_1(x)$
and there are no other fixed points with $0 < x < 1$.
The coexistence of a giant component and 
non-trivial finite tree components (that consist of more than a single node) appears only
in case that the degree distributions $P(k)$ supports a non-trivial solution of Eq. (\ref{eq:tg}),
in which $0 < \tilde g < 1$.
This requires a non-zero probability of leaf-nodes, namely $P(1)>0$,
and thus occurs only when $k_{\rm min}=0$ or $k_{\rm min}=1$.
In large configuration model networks in which $k_{\rm min} \ge 2$ 
and the mean degree satisfies the condition $c > 2$, the
giant component encompasses the whole network and $g = \tilde g = 1$
\cite{Bonneau2017}.

Here we focus on configuration model networks with 
degree distributions $P(k)$, which are bounded from below by
$k_{\rm min}=0$ or $1$.
Under suitable conditions, such networks may 
exhibit a coexistence between a giant component and  
finite tree components.
The condition for the existence of a giant component can be expressed in
the form
\begin{equation}
\frac{ \langle K^2 \rangle }{ \langle K \rangle }  - 1 > 1,
\label{eq:ML}
\end{equation}

\noindent
which is known as the Molloy-Reed criterion
\cite{Molloy1995,Molloy1998}.
In order to discuss this condition, consider a node $i$
that is sampled via a random edge $e$.
The excess degree $k_{\rm ex}$ of $i$ is the number of 
other edges apart from the edge $e$, namely $k_{\rm ex}=k-1$,
where $k$ is the degree of $i$.
In essence, the condition of Eq. (\ref{eq:ML})
states that a giant component exists if the expectation
value of the excess degree 
of nodes sampled via a random edge exceeds $1$.
Thus, the percolation threshold $c_0$ is the value of 
the mean degree
$\langle K \rangle$
at which   
$ \langle K^2 \rangle = 2 \langle K \rangle$.

\section{Statistical analysis of nodes}

Below we analyze the statistical properties of randomly sampled nodes in
configuration model networks. We calculate the probability that a random node resides on
the giant component (and the complementary probability that it resides on
one of the finite components). We also analyze the distinct statistical properties
of the nodes that reside on the giant component and on the finite components.


\begin{figure}[t!]
\includegraphics[width=5cm]{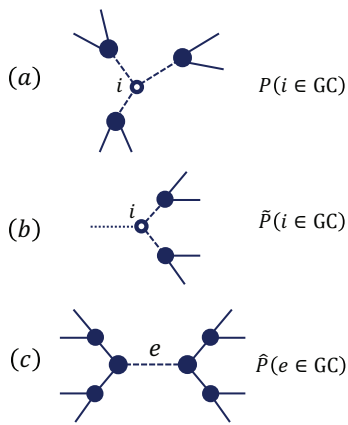} 
\caption{
(a) A random node $i$ (empty circle) of degree $k$ in a configuration model network (left).
The probability that $i$ does not reside on the giant component is equal to the probability 
that none of its $k$ neighbors (full circles) resides on the giant component of the reduced
network (right) from which $i$ is removed, together with its links (dashed lines).
(b) A node $i$  (empty circle) of degree $k$ sampled via a random edge (left),
which is marked as a dashed line. We are interested in the probability that $i$ does not reside on the giant 
component of the reduced network from which the sampled edge (dashed line) is removed.
This probability is equal to the probability that none of its $k-1$ remaining neighbors of $i$ resides on
the giant component of
the further reduced network (right) from which the node $i$ is removed together with its links
(dashed lines).
(c) A random edge $e$ with end-nodes $i$ and $i'$ of degrees $k$ and $k'$, respectively (left).
The probability that $e$ does not reside on the giant component is equal to the probability
that none of its two end-nodes resides on the giant component of the reduced network  
from which $e$ is removed.
This probability is equal to the probability that none of the $k-1$ remaining neighbors of $i$
and none of the $k'-1$ remaining neighbors of $i'$ resides on the giant component of the further 
reduced network (right) from which $i$ and $i'$ are removed together with their links
(dashed lines).
}
\label{fig:2}
\end{figure}

\subsection{The fraction of nodes that reside on the giant/finite components}

The probability that a random node $i$ in a configuration model network
resides on the giant component is 
\cite{Tishby2018b,Tishby2019}
\begin{equation}
P(i \in {\rm GC}) = g,
\label{eq:PiGC}
\end{equation}

\noindent
where $g$ is given by Eq. (\ref{eq:g}),
while the probability that it resides on one of the finite components is
\begin{equation}
P(i \in {\rm FC}) = 1 - g.
\label{eq:PiFC}
\end{equation}

\noindent
A node $i$ of a given degree $k$ resides on the giant component if at least one of its $k$ neighbors
resides on the giant component of the reduced network from which $i$ is removed
[Fig. 2(a)].
Using the theoretical framework of the cavity method
\cite{Mezard1985,Mezard2003,Mezard2009,Ferraro2015},
each neighbor of $i$ can be considered as a node selected via a random edge.
Therefore, the probability that each one of the neighbors of $i$
resides on the giant component of the reduced network 
from which $i$ is removed
is given by $\tilde g$.
Moreover, 
due to the locally tree-like structure of configuration model networks,
the probabilities of different neighbors of $i$ to reside on the giant 
component of the reduced network 
from which $i$ is removed
are independent of each other.
Therefore, the probability that a node $i$
selected randomly from all the nodes of degree $k$
in the network
resides on the giant component,
is given by
\cite{Tishby2018b,Tishby2019}
\begin{equation}
P(i \in {\rm GC} | k)= 1 - (1 - \tilde g)^k,
\label{eq:PiGCk}
\end{equation}

\noindent
where $\tilde g$ is given by Eq. (\ref{eq:tg}),
while the probability that it resides on one of the finite tree components is
given by
\begin{equation}
P(i \in {\rm FC} | k) =  (1 - \tilde g)^k.
\label{eq:PiFCk}
\end{equation}


\begin{figure}[ht!]
\includegraphics[width=7.5cm]{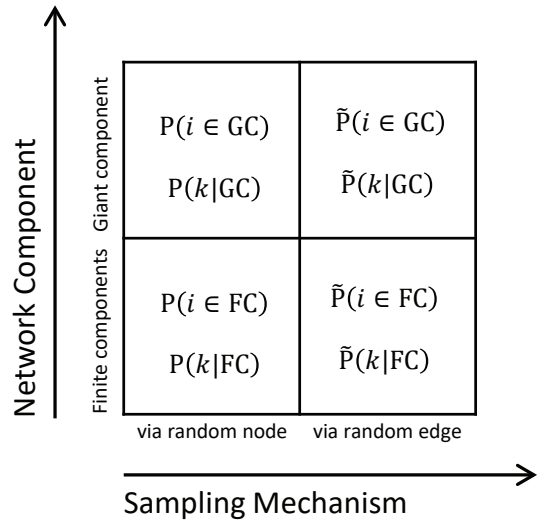} 
\caption{
Illustration of the four categories of nodes considered in this paper, presented
in the form of a two by two matrix diagram.
The horizontal axis accounts for the two sampling procedures,
namely random node sampling and node sampling via random edges.
The vertical axis accounts for the location of a node in the network,
which can be either on the giant component or on one of the finite tree components.
Each one of the four categories of nodes exhibits different statistical properties.
}
\label{fig:3}
\end{figure}

\noindent
Clearly, the probability of $i$ to reside on the giant component is an increasing function
of the degree $k$, while the probability of $i$ to reside on one of the finite components
is a decreasing function of $k$.

The different categories of nodes 
in configuration model networks,
in terms of the sampling procedure and their
location in the network,
are illustrated in Fig. 3
in the form of a two by two matrix diagram.
The horizontal axis accounts for the two sampling procedures,
namely random node sampling and node sampling via random edges.
The vertical axis accounts for the location of a node in the network,
which can be either on the giant component or on one of the finite tree components.
Each one of the four categories of nodes exhibits different statistical properties.
Such $2 \times 2$ matrix diagrams are used extensively in the analysis of decision making
processes and business management
\cite{Lowy2004}.

\subsection{The degree distributions of nodes on the giant/finite components}

The micro-structure of the giant component of configuration model
networks was recently studied
\cite{Tishby2018b,Tishby2019}.
It was shown that the degree distribution,
conditioned on the giant component, is given by
\begin{equation}
P(k | {\rm GC}) =
\frac{ 1- (1-\tilde g)^k}{g} P(k),
\label{eq:PkGC}
\end{equation}

\noindent
while the degree distribution
conditioned on the finite components
is given by
\begin{equation}
P(k | {\rm FC}) =
\frac{ (1-\tilde g)^k}{1 - g} P(k),
\label{eq:PkFC}
\end{equation}

\noindent
where $k \ge k_{\rm min}$.
In the analysis below we focus on degree distributions whose support
is bounded from below by either $k_{\rm min}=0$ or $k_{\rm min}=1$,
which enable the coexistence between the giant component and the finite tree components.
The derivations apply to both cases.
The specific value of $k_{\rm min}$ is not specified in each equation, but it is implicitly assumed that
in the case of $k_{\rm min}=1$ the probability $P(0)=0$.

As expected, Eq. (\ref{eq:PkGC}) satisfies $P(0|{\rm GC})=0$
even for $k_{\rm min}=0$, namely there are
no isolated nodes on the giant component. 
Isolated nodes are considered as finite tree components of size $s=1$.
The probability that a random node on the finite components
is an isolated node is given by $P(0|{\rm FC})=P(0)/(1-g)$,
namely the fraction of isolated nodes on the finite components is higher than in the whole network.
Regarding leaf nodes of degree $k=1$, their fraction on the giant 
component,
given by $P(1|{\rm GC}) = (\tilde g/g) P(1)$,
is higher than in the whole network in case that $g < \tilde g$
and lower in case that $g > \tilde g$. 
Since $g > \tilde g[1-P(0)]$ the former case may occur only in networks that include 
isolated nodes, in which $P(0)>0$
\cite{Tishby2018a}.
Since 
in the coexistence phase, where $0 < g,\tilde g < 1$,
the probabilities $g$ and $\tilde g$ satisfy the condition
\cite{Tishby2018a}
\begin{equation}
\frac{ (1-\tilde g)^2 }{ 1-g } < 1,
\label{eq:gtg1}
\end{equation}

\noindent
the fraction of nodes of degrees $k \ge 2$
on the finite tree components is lower than in the whole network.
The condition of Eq. (\ref{eq:gtg1}) can also be expressed in the form
\begin{equation}
\frac{ (2-\tilde g) \tilde g}{ g } > 1.
\label{eq:gtg2}
\end{equation}

\noindent
Note that the numerator on the left hand side of Eq. (\ref{eq:gtg2})
satisfies 
\begin{equation}
(2-\tilde g)\tilde g < 1.
\label{eq:2g2s1}
\end{equation}

\noindent
To show this we define $\tilde h = 1 - \tilde g$
and obtain
$(2-\tilde g)\tilde g= 1-\tilde h^2 < 1$.
The degree distribution of the whole network is recovered by
\begin{equation}
P(k) = P(k|{\rm GC}) P(i \in {\rm GC}) + P(k|{\rm FC}) P(i \in {\rm FC}),
\label{eq:PkGCFC}
\end{equation}

\noindent
where $P(i \in {\rm GC})$ and $P(i \in {\rm FC})$
are given by Eqs. (\ref{eq:PiGC}) and (\ref{eq:PiFC}), respectively.

The giant component of a configuration model network consists of a 2-core
which is decorated by tree branches.
The 2-core (2-CORE) is a connected component, such that
each node on the 2-core has links to at least two other nodes
that reside on the 2-core
\cite{Seidman1983,Dorogovtsev2006a,Dorogovtsev2006b,Yuan2016}.
Moreover, each node on the 2-core 
of a configuration model network
resides on at least one cycle.
The nodes on the tree branches belong to the 1-core of the giant
component but not to the 2-core.
This is expressed by $i \in {\rm GC} \cap \overline{\mbox{2-CORE}}$,
where $\overline{X}$ represents the complementary set of $X$
and $X \cap Y$ is the intersection of $X$ and $Y$.
The degree distribution of the nodes on the 2-core of the giant
component is given by
\begin{equation}
P(k |    \mbox{2-CORE}   ) =
\frac{ 1 - (1-\tilde g)^k - k \tilde g (1-\tilde g)^{k-1}   }{g - \tilde g (1-\tilde g) \langle K \rangle} P(k),
\label{eq:PkGC2C}
\end{equation}

\noindent
while the degree distribution of the nodes on the tree branches of the giant component
is given by
\begin{equation}
P(k | {\rm GC} \cap   \overline{\mbox{2-CORE}}   ) =
(1-\tilde g)^{k-2}  \widetilde P(k).
\label{eq:PkGCN2C}
\end{equation}

\noindent
The probability that a random node on the giant component
resides on the 2-core is given by
\begin{equation}
P(i \in    \mbox{2-CORE}  | {\rm GC}) = 1 - \frac{ \tilde g (1-\tilde g) }{g} \langle K \rangle,
\end{equation}

\noindent
while the probability that it resides on one of the tree branches
is given by
\begin{equation}
P(i \in   \overline{\mbox{2-CORE}}  |{\rm GC}) = \frac{ \tilde g (1-\tilde g) }{g} \langle K \rangle.
\end{equation}

\subsection{The mean degrees of nodes on the giant/finite components}

The mean degree of the nodes that reside on the giant component is
given by
%
$\mathbb{E}[ K | {\rm GC} ] = \sum_{k} k P(k|{\rm GC})$.
%
Inserting $P(k|{\rm GC})$ from Eq. (\ref{eq:PkGC}) 
and carrying out the summation, we obtain
\begin{equation}
\mathbb{E}[K|{\rm GC}] 
= 
\frac{  (2-\tilde g) \tilde g }{g} 
\langle K \rangle.
\label{eq:EKGC}
\end{equation}

\noindent
Using Eq. (\ref{eq:gtg2}) we conclude that
$\mathbb{E}[K|{\rm GC}] > \langle K \rangle$,
namely the mean degree of the nodes that reside on the giant component is
larger than the mean degree of the whole network.

The mean degree of the nodes that reside on the finite tree components
is denoted by $\mathbb{E}[K|{\rm FC}]$.
Using $P(k|{\rm FC})$ from Eq. (\ref{eq:PkFC}),
we obtain
\begin{equation}
\mathbb{E}[K|{\rm FC}]
= \frac{ (1-\tilde g)^2 }{1-g} \langle K \rangle.
\label{eq:EKFC}
\end{equation}

\noindent
Using Eq. (\ref{eq:gtg1}) we conclude that the mean degree of the nodes
that reside on the finite tree components is smaller than the mean degree
of the whole network, namely $\mathbb{E}[k|{\rm FC}] < \langle K \rangle$.
The mean degree of the whole network is recovered by
\begin{equation}
\langle K \rangle = {\mathbb E}[K|{\rm GC}] P(i \in {\rm GC}) + {\mathbb E}[K|{\rm FC}] P(i \in {\rm FC}),
\label{eq:EKGCFC}
\end{equation}

\noindent
where $P(i \in {\rm GC})$ and $P(i \in {\rm FC})$
are given by Eqs. (\ref{eq:PiGC}) and (\ref{eq:PiFC}), respectively.

\subsection{The variance of the degree distributions on the giant/finite components}

The second moment 
of the degree distribution
$P(k|{\rm GC})$
of the nodes that reside on the giant component is
given by
%
%
%
\begin{equation}
\mathbb{E}[K^2|{\rm GC}] 
= 
\frac{1}{g}
\biggl\{
\langle K^2 \rangle  
-   (1-\tilde g)^2   \big[ 1 + G'_1(1-\tilde g) \big] \langle K \rangle
\biggr\},
\label{eq:EK2GC}
\end{equation}

\noindent
where $G'_1(x)$ is the derivative of $G_1(x)$.
Since the fixed point of Eq. (\ref{eq:tg}) is a stable fixed point,
the derivative satisfies
$G'_1(1-\tilde g) < 1$.
Writing $G'_1(1-\tilde g)$ explicitly, in the form
\begin{equation}
G'_1(1-\tilde g) = \frac{1}{\langle K \rangle}
\sum_{k=2}^{\infty} k(k-1) (1-\tilde g)^{k-2} P(k),
\label{eq:G1'tg}
\end{equation}

\noindent 
we find that it satisfies
\begin{equation}
G'_1(1-\tilde g)  < \min \left\{  \frac{ \langle K^2 \rangle }{\langle K \rangle} - 1, 1   \right\}.
\label{eq:G1'}
\end{equation}

\noindent
Inserting this result into Eq. (\ref{eq:EK2GC}) and using
Eq. (\ref{eq:gtg2}), it is found that
in the coexistence phase, where $0 < g, \tilde g < 1$,
$\mathbb{E}[K^2|{\rm GC}] > \langle K^2 \rangle$.
In the dilute network regime of $0 < \tilde g \ll 1$,
just above the percolation transition,
one can expand the right hand side of Eq. (\ref{eq:G1'tg})
to first order in $\tilde g$ and obtain
\begin{equation}
G'_1(1-\tilde g) 
\simeq  \frac{ \langle K^2 \rangle }{\langle K \rangle} -1
- \left( \frac{ \langle K^3 \rangle }{\langle K \rangle}
- 3 \frac{ \langle K^2 \rangle }{\langle K \rangle} + 2 \right) \tilde g
+ \mathcal{O} \left(  \tilde g^2  \right).
\label{eq:G1'b}
\end{equation}

\noindent
The variance of
$P(k|{\rm GC})$ 
is given by
%
%
%
\begin{eqnarray}
{\mathbb V}[K|{\rm GC}] &=&  
\frac{1}{g}
\biggl\{
\langle K^2 \rangle  
-   (1-\tilde g)^2   \big[ 1 + G'_1(1-\tilde g) \big] \langle K \rangle
\biggr\}\nonumber\\
& &-
\frac{  [ (2-\tilde g) \tilde g ]^2 }{g^2} 
\langle K \rangle^2.
\label{eq:VKGC}
\end{eqnarray}

\noindent
While both the first and second moments of $P(k|{\rm GC})$ are larger than the 
corresponding moments of $P(k)$, the variance ${\mathbb V}[K|{\rm GC}]$ may be
either larger or smaller than ${\mathbb V}[K]$, depending on the specific properties
of the degree distribution.

The second moment 
of the degree distribution
$P(k|{\rm FC})$
of the nodes that reside on the finite tree components is
denoted by $\mathbb{E}[ K^2 | {\rm FC} ]$.
Using $P(k|{\rm FC})$ from Eq. (\ref{eq:PkFC}), 
we obtain
\begin{equation}
\mathbb{E}[K^2|{\rm FC}] 
= 
\frac{ (1-\tilde g)^2 }{1-g} \left[ 1 + G'_1(1-\tilde g) \right] \langle K \rangle.
\label{eq:EK2FC}
\end{equation}

\noindent
Using Eqs. (\ref{eq:gtg1}) and (\ref{eq:G1'})
one can show that 
$\mathbb{E}[K^2|{\rm FC}] < \langle K^2 \rangle$.
The variance of 
$P(k|{\rm FC})$ 
is denoted by ${\mathbb V}[K|{\rm FC}]$.
Using the first and second moments from Eqs. (\ref{eq:EKFC}) and (\ref{eq:EK2FC}),
respectively, we obtain
\begin{equation}
{\mathbb V}[K|{\rm FC}] =  
\frac{ (1-\tilde g)^2 }{1-g} 
\langle K \rangle
\left\{
1 + G'_1(1-\tilde g)  
-
\frac{ (1-\tilde g)^2 }{ 1-g } \langle K \rangle 
\right\}.
\label{eq:VKFC}
\end{equation}

\noindent
While both the first and second moments of $P(k|{\rm FC})$ are smaller than the 
corresponding moments of $P(k)$, the variance ${\mathbb V}[K|{\rm FC}]$ may be
either larger or smaller than ${\mathbb V}[K]$, depending on the specific properties
of the degree distribution.


\begin{figure}[ht!]
\includegraphics[width=7.5cm]{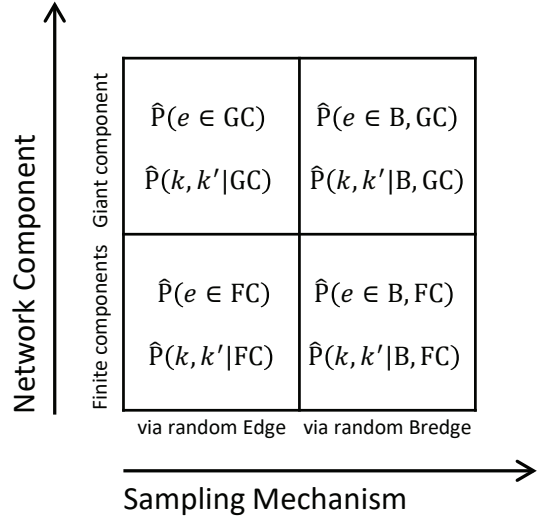} 
\caption{
Illustration of the four categories of edges considered in this paper, presented
in the form of a two by two matrix diagram.
The horizontal axis accounts for the two sampling procedures,
namely random sampling from all the edges in the network 
or random sampling restricted to those edges which are bredges.
The vertical axis accounts for the location of an edge in the network,
which can be either on the giant component or in one of the finite tree components.
Each one of the four categories of edges exhibits different statistical properties.
}
\label{fig:4}
\end{figure}

\section{Statistical analysis of edges}

Below we analyze the statistical properties of randomly selected edges in
configuration model networks. We calculate the probability that a random edge resides on
the giant component (and the complementary probability that it resides on
one of the finite components). We also analyze the distinct statistical properties
of the edges that reside on the giant component and 
of those that reside on the finite components.

The different categories of edges in terms of the sampling procedure and their
location in the network are illustrated in Fig. 4.
The horizontal axis accounts for the two sampling procedures,
namely random sampling from all the edges in the network 
or random sampling restricted to those edges which are bredges.
The vertical axis accounts for the location of an edge in the network,
which can be either on the giant component or in one of the finite tree components.
Each one of the four categories of edges exhibits different statistical properties.

\subsection{The fraction of edges that reside on the giant/finite components}

Consider a randomly chosen end-node $i$ of a random edge $e$ [Fig. 2(b)].
The probability that $i$ resides
on the giant component of the reduced network from which $e$ is removed
is 
\begin{equation}
\widetilde P(i \in {\rm GC}) = \tilde g,
\label{eq:tPiGC}
\end{equation}

\noindent
where $\tilde g$ is given by Eq. (\ref{eq:tg}),
while the probability that $i$ resides on one of the finite components
of the reduced network is
\begin{equation}
\widetilde P(i \in {\rm FC}) = 1 - \tilde g.
\label{eq:tPiFC}
\end{equation}

Consider a random edge $e$ [Fig. 2(c)].
The probability that $e$ resides on one of the finite tree components
of the network
amounts to the probability that both its end-nodes reside on finite components
of the reduced network from which $e$ is removed.
It is thus given by
\begin{equation}
\widehat P(e \in {\rm FC}) =  (1 - \tilde g)^2.
\label{eq:hPeFC}
\end{equation}

\noindent
Therefore, the complementary probability that a random edge $e$ resides on the giant component is
\begin{equation}
\widehat P(e \in {\rm GC}) =  (2-\tilde g) \tilde g.
\label{eq:hPeGC}
\end{equation}

\noindent
The degrees of end-nodes satisfy $k \ge 1$ even in case that $k_{\rm min}=0$.
The probability that the end-node $i$ belongs to the giant component of the reduced network 
from which $e$ is removed,
is given by
\begin{equation}
\widetilde P(i \in {\rm GC}| k) = 1 - (1-\tilde g)^{k-1},
\label{eq:tPiGCk}
\end{equation}

\noindent
while the probability that it belongs to one of the finite components of the reduced network is
\begin{equation}
\widetilde P(i \in {\rm FC}| k) = (1-\tilde g)^{k-1},
\label{eq:tPiFCk}
\end{equation} 

\noindent
where $k \ge 1$.

Consider a random edge $e$ whose end-nodes $i$ and $i'$ 
are of degrees $k \ge 1$ and $k' \ge 1$, respectively.
The probability that such an edge resides on the giant component
is given by
\begin{equation}
\widehat P(e \in {\rm GC|k,k'}) = 1 - (1-\tilde g)^{k+k'-2},
\label{eq:hPeGCkk'}
\end{equation}

\noindent
while the probability that it resides on one of the finite tree components 
is
\begin{equation}
\widehat P(e \in {\rm FC|k,k'}) = (1-\tilde g)^{k+k'-2}.
\label{eq:hPeFCkk'}
\end{equation}

\noindent
Interestingly, these probabilities depend only on the sum of $k$ and $k'$
rather than on each one of them separately.
For $k=k'=1$ one obtains
$\widehat P(e \in {\rm GC|1,1})=0$
and
$\widehat P(e \in {\rm FC|1,1})=1$.
This is due to the fact that in this case $i$ and $i'$ 
form a dimer, which is isolated from the rest of the network.
As the sum $k+k'$ increases, the probability
that the edge $e$ resides on one of the finite components
decays exponentially while the probability that it resides on
the giant component converges towards $1$.

\subsection{The marginal degree distributions of end-nodes}

The degree distribution of the end-nodes of random edges 
is given by Eq. (\ref{eq:tPk}),
where $k \ge 1$.
The degree distribution of the end-nodes of random edges
that reside on the giant component is given by
\begin{equation}
\widetilde P(k|{\rm GC}) = \frac{k}{\mathbb{E}[K|{\rm GC}]} P(k|{\rm GC}).
\label{eq:tPkGCd}
\end{equation}

\noindent
Inserting $P(k|{\rm GC})$ from Eq. (\ref{eq:PkGC}) 
and $\mathbb{E}[K|{\rm GC}]$ from Eq. (\ref{eq:EKGC}), 
we obtain
\begin{equation}
\widetilde P(k|{\rm GC}) =  \frac{ 1 - (1-\tilde g)^{k} }{(2-\tilde g) \tilde g} \widetilde P(k),
\label{eq:tPkGC}
\end{equation}

\noindent
where $k \ge 1$.
From Eq. (\ref{eq:tPkGC}) one finds that the fraction of
end-nodes of degree $k=1$ on the giant component,
given by $\widetilde P(1|{\rm GC}) = \widetilde P(1)/(2-\tilde g)$, 
is lower than in the whole network.
Interestingly, the fraction of end-nodes of degree $k=2$ on the giant component,
given by $\widetilde P(2|{\rm GC})=\widetilde P(2)$,
is identical to their fraction in the whole network.
For $k \ge 3$ it is found that
$\widetilde P(k|{\rm GC}) > \widetilde P(k)$,
namely nodes of degrees $k \ge 3$ are more probable
on the giant component compared to the whole network.
In the limit of $k \rightarrow \infty$
$\widetilde P(k|{\rm GC}) \rightarrow \widetilde P(k)/[(2-\tilde g)\tilde g]$,
where
$(2-\tilde g)\tilde g < 1$ [Eq. (\ref{eq:2g2s1})].
The degree distribution of the end-nodes of random edges that
reside on the finite components 
is given by
\begin{equation}
\widetilde P(k|{\rm FC}) = \frac{k}{\mathbb{E}[K|{\rm FC}]} P(k|{\rm FC}).
\label{eq:tPkFCd}
\end{equation}

\noindent
Inserting $P(k|{\rm FC})$ from Eq. (\ref{eq:PkFC}) 
and $\mathbb{E}[K|{\rm FC}]$ from Eq. (\ref{eq:EKFC}), 
we obtain
\begin{equation}
\widetilde P(k|{\rm FC}) = (1-\tilde g)^{k-2} \widetilde P(k),
\label{eq:tPkFC}
\end{equation}

\noindent
where $k \ge 1$.
The degree distribution $\widetilde P(k)$ of the end-nodes of random
edges in the network is recovered by
\begin{equation}
\widetilde P(k) = 
\widetilde P(k|{\rm GC})   \widehat P(e \in {\rm GC}) 
+
\widetilde P(k|{\rm FC})   \widehat P(e \in {\rm FC}),
\end{equation}

\noindent
where $\widetilde P(k|{\rm GC})$ is given by Eq. (\ref{eq:tPkGC}),
$\widetilde P(k|{\rm FC})$ is given by Eq. (\ref{eq:tPkFC}),
$\widehat P(e \in {\rm GC})$ is given by Eq. (\ref{eq:hPeGC})
and $\widehat P(e \in {\rm FC})$ is given by Eq. (\ref{eq:hPeFC}).

From Eq. (\ref{eq:tPkFC}) one finds that the fraction of
end-nodes of degree $k=1$ on
the finite components,
given by 
$\widetilde P(1|{\rm FC}) = \widetilde P(1)/(1-\tilde g)$,
is higher than in the whole network.
The fraction of end-nodes of degree $k=2$ on the finite components
is identical to their fraction in the whole network.
For any value of $k \ge 3$ the fraction of end-nodes of degree $k$ on the finite components is 
lower than in the whole network.
The 'phase separation' between the giant component and the finite components
may thus be considered as a distillation process, in which high-degree nodes
tend to concentrate on the giant component
while low-degree nodes end up in the finite components.

\subsection{The mean degrees of end-nodes}

The mean degree of end-nodes of random edges is denoted by $\widetilde{ \mathbb{E}}[K]$.
Using $\widetilde P(k)$ from Eq. (\ref{eq:tPk}),
we obtain
\begin{equation}
\widetilde{ \mathbb{E}}[K]  
= 
\frac{ \langle K^2 \rangle }{\langle K \rangle}.
\label{eq:tEK}
\end{equation}

\noindent
The mean degree of end-nodes of random edges that reside on  
the finite tree components is denoted by
$\widetilde{ \mathbb{E}}[K|{\rm FC}]$.
Using $\widetilde P(k|{\rm FC})$ from Eq. (\ref{eq:tPkFC}),
we obtain
\begin{equation}
\widetilde{ \mathbb{E}}[K|{\rm FC}]
=
1 + G_1'(1-\tilde g),
\label{eq:tEKFC}
\end{equation}

\noindent
where
$G'_1(x)$ is the derivative of $G_1(x)$.
Interestingly, this implies that
$G_1'(1-\tilde g)$ can be interpreted as the mean 
excess degree 
$\widetilde{ \mathbb{E}}[K_{\rm ex}|{\rm FC}]  = \widetilde{ \mathbb{E}}[K|{\rm FC}] - 1$
of end-nodes of edges that reside on the finite components.
In general, the mean excess degree of nodes sampled via random edges is analogous to the
basic reproduction ratio $R_0$ of infectious diseases
\cite{Diekmann2000}
and to the neutron multiplication factor of nuclear chain reactions
\cite{Duderstadt1976}.
Using Eq. (\ref{eq:G1'}) it is found that
$\widetilde{ \mathbb{E}}[K|{\rm FC}] < \widetilde{ \mathbb{E}}[K]$.

The mean degree of end-nodes of random edges that reside on  
the giant component is denoted by
$\widetilde{ \mathbb{E}}[K|{\rm GC}]$.
Using $\widetilde P(k|{\rm GC})$ from Eq. (\ref{eq:tPkGC}),
we obtain
\begin{equation}
\widetilde{ \mathbb{E}}[K|{\rm GC}] =
\frac{1}{(2-\tilde g)\tilde g}
\left\{
\frac{ \langle K^2 \rangle }{ \langle K \rangle }
-
(1-\tilde g)^2 [1 + G_1'(1-\tilde g)]
\right\}.
\label{eq:tEKGC}
\end{equation}

\noindent
Using Eq. (\ref{eq:G1'}) it is found that
$\widetilde{ \mathbb{E}}[K|{\rm GC}] > \widetilde{ \mathbb{E}}[K]$.
Note that in heavy tail degree distributions
the mean degree 
$\widetilde{ \mathbb{E}}[K|{\rm GC}]$ 
of the end-nodes that reside on the giant component
may diverge even under conditions in which
$\langle K \rangle$ is finite.
This is due to the fact that in such distributions the second moment
$\langle K^2 \rangle$ that appears on the right hand side
of Eq. (\ref{eq:tEKGC})
may diverge, leading to the divergence of
$\widetilde{ \mathbb{E}}[K|{\rm GC}]$.
The mean degree $\widetilde{\mathbb{E}}[K]$ can be recovered by
\begin{equation}
\widetilde{ \mathbb{E}}[K]  
=
\widetilde{ \mathbb{E}}[K|{\rm GC}]      \widehat P(e \in {\rm GC})
+
\widetilde{ \mathbb{E}}[K|{\rm FC}]     \widehat P(e \in {\rm FC}).
\label{eq:tEKGCFC}
\end{equation}

\subsection{The variance of the degree distribution of end-nodes}

The second moment of 
the degree distribution
$\widetilde P(k)$ 
of the end-nodes of random edges
is denoted by
$\widetilde{ \mathbb{E}}[K^2]$.
Using $\widetilde P(k)$ from Eq. (\ref{eq:tPk}),
we obtain
\begin{equation}
\widetilde{ \mathbb{E}}[K^2]
= 
\frac{ \langle K^3 \rangle }{\langle K \rangle}.
\label{eq:tEK2}
\end{equation}

\noindent
The variance of $\widetilde P(k)$ is denoted by
$\widetilde {\mathbb V}[K]$.
Using the first moment
$\widetilde{ \mathbb{E}}[K]$
from Eq. (\ref{eq:tEK})
and 
the second moment
$\widetilde{ \mathbb{E}}[K^2]$
from Eq. (\ref{eq:tEK2}),
we obtain
\begin{equation}
\widetilde {\mathbb V}[K] = 
\frac{ \langle K^3 \rangle }{\langle K \rangle}
-
\frac{ \langle K^2 \rangle^2 }{ \langle K \rangle^2 }.
\label{eq:tVK2}
\end{equation}

The second moment of the degree distribution
$\widetilde P(k|{\rm FC})$
of end-nodes of random edges that reside on  
the finite tree components is denoted by
$\widetilde{\mathbb{E}}[K^2|{\rm FC}]$.
Using $\widetilde P(k|{\rm FC})$ from Eq. (\ref{eq:tPkFC}),
we obtain
\begin{equation}
\widetilde{\mathbb{E}}[K^2|{\rm FC}]
=
(1-\tilde g) G''_1(1-\tilde g) 
+
3 G_1'(1-\tilde g) + 1,
\label{eq:tEK2FC}
\end{equation}

\noindent
where $G''_1(x)$ is the second derivative of $G_1(x)$.
Writing $G''_1(1-\tilde g)$ explicitly in the form
\begin{equation}
G''_1(1-\tilde g) = \frac{1}{\langle K \rangle}
\sum_{k=3}^{\infty}
k(k-1)(k-2) (1-\tilde g)^{k-3} P(k),
\label{eq:G1''}
\end{equation}

\noindent
we find that it satisfies
\begin{equation}
G''_1(1-\tilde g) \le
\frac{\langle K^3 \rangle}{\langle K \rangle}
-3 \frac{\langle K^2 \rangle}{\langle K \rangle}
+ 2,
\label{eq:G1''ts}
\end{equation}

\noindent
where equality is obtained for $\tilde g=0$.
Combining this result with Eq. (\ref{eq:G1'}), it is found that
in the coexistence phase, where $0 < g,\tilde g < 1$,
the second moment satisfies
$\widetilde{\mathbb{E}}[K^2|{\rm FC}] < \widetilde{\mathbb{E}}[K^2]$.
In the dilute network regime of $0 < \tilde g \ll 1$,
just above the percolation transition,
one can expand the right hand side of Eq. (\ref{eq:G1''})
to first order in $\tilde g$ and obtain
\begin{eqnarray}
G''_1(1-\tilde g) & \simeq &
\frac{\langle K^3 \rangle}{\langle K \rangle}
-3 \frac{\langle K^2 \rangle}{\langle K \rangle}
+ 2 - \Bigg( \frac{\langle K^4 \rangle}{\langle K \rangle}
-6 \frac{\langle K^3 \rangle}{\langle K \rangle}
\nonumber\\
& &+ 11 \frac{\langle K^2 \rangle}{\langle K \rangle}
- 6 \Bigg) \tilde g + \mathcal{O} \left( \tilde g^2\right).
\label{eq:G1''t}
\end{eqnarray}

\noindent
Using Eq. (\ref{eq:G1''ts})
it is found that
$\widetilde{\mathbb{E}}[K^2|{\rm FC}] < \widetilde{\mathbb{E}}[K^2]$.

The variance of the degree distribution $\widetilde P(k|{\rm FC})$ of the end-nodes that reside
on the finite components is denoted by
$\widetilde{\mathbb{V}}[K|{\rm FC}]$.
Using the first and second moments from Eqs. (\ref{eq:tEKFC}) and (\ref{eq:tEK2FC}),
respectively, we obtain
\begin{equation}
\widetilde{\mathbb{V}}[K|{\rm FC}] =  
(1-\tilde g) G''_1(1-\tilde g) 
+
G_1'(1-\tilde g)  \big[ 1 - G_1'(1-\tilde g) \big].
\label{eq:tVKFC}
\end{equation}

\noindent
While both the first and second moments of $\widetilde P(k|{\rm FC})$ are smaller than the 
corresponding moments of $\widetilde P(k)$, the variance $\widetilde{{\mathbb V}}[K|{\rm FC}]$ may be
either larger or smaller than $\widetilde{{\mathbb V}}[K]$, depending on the specific properties
of the degree distribution.

The second moment of the degree distribution
$\widetilde P(k|{\rm GC})$
of the end-nodes that reside on  
the giant component is denoted by
$\widetilde{\mathbb{E}}[K^2|{\rm GC}]$.
Using $\widetilde P(k|{\rm GC})$ from Eq. (\ref{eq:tPkGC}),
we obtain
\begin{eqnarray}
\widetilde{\mathbb{E}}[K^2|{\rm GC}]
& =&
\frac{1}{ (2-\tilde g) \tilde g }
\Bigg\{
\frac{ \langle K^3 \rangle }{ \langle K \rangle} - (1-\tilde g)^2 
\big[ (1-\tilde g)  \nonumber\\
& & 
\times G''_1(1-\tilde g) 
+
3 G_1'(1-\tilde g) + 1 \big]
\Bigg\}.
\label{eq:tEK2GC}
\end{eqnarray}

\noindent
Using Eq. (\ref{eq:G1''ts})
it is found that
$\widetilde{\mathbb{E}}[K^2|{\rm GC}] > \widetilde{\mathbb{E}}[K^2]$.
The variance of the degree distribution $\widetilde P(k|{\rm GC})$ of the end-nodes that reside
on the giant component is denoted by
$\widetilde{\mathbb{V}}[ K|{\rm GC}]$.
Using the first and second moments from Eqs. (\ref{eq:tEKGC}) and (\ref{eq:tEK2GC}),
respectively, we obtain
\begin{eqnarray}
\widetilde{\mathbb{V}}[K|{\rm GC}] 
&=&  
\frac{1}{ (2-\tilde g) \tilde g }
\Bigg\{
\frac{\langle K^3 \rangle }{ \langle K \rangle}
- 
(1-\tilde g)^2 \big[ (1-\tilde g) \nonumber\\
& & \times G''_1(1-\tilde g) 
+
3 G_1'(1-\tilde g) + 1 \big]
\Bigg\}
\nonumber \\
&-&\hspace{-7mm}
\frac{1}{[ (2-\tilde g) \tilde g]^2}
\left\{ 
\frac{ \langle K^2 \rangle }{ \langle K \rangle} 
- (1-\tilde g) \big[ 1 + G_1'(1-\tilde g) \big]  
\right\}^2.\nonumber\\
\label{eq:tVKGC}
\end{eqnarray}

\noindent
While both the first and second moments of $\widetilde P(k|{\rm GC})$ are larger than the 
corresponding moments of $\widetilde P(k)$, the variance $\widetilde{{\mathbb V}}[K|{\rm GC}]$ may be
either larger or smaller than $\widetilde{{\mathbb V}}[K]$, depending on the specific properties
of the degree distribution.
The variance 
$\widetilde{\mathbb{V}}[K|{\rm GC}]$
is used below as a normalization factor for the 
covariance of the joint degree distribution
of edges that reside on the giant component,
which yields the Pearson correlation coefficient.

\subsection{The joint degree distribution of end-nodes}

The joint degree distribution 
$\widehat P(k,k')$ 
of the end-nodes $i$ and $i'$ of a random edge 
in a configuration model network with degree distribution $P(k)$ is given by
\begin{equation}
\widehat P(k,k') = \widetilde P(k) \widetilde P(k'),
\label{eq:hPkk'}
\end{equation}

\noindent
where $\widetilde P(k)$ is given by Eq. (\ref{eq:tPk}).
Note that in Eq. (\ref{eq:hPkk'})
the degrees satisfy
$k,k' \ge 1$.
The end-nodes $i$ and $i'$ are considered as two distinguishable objects.
Thus, $\widehat P(k,k')$ is the probability that $i$ is of degree $k$ and
$i'$ is of degree $k'$.
The probability that $i$ is of degree $k'$ and $i'$ is of degree $k$ is 
given by
$\widehat P(k',k) = \widehat P(k,k')$.
Therefore, the probabilities $\widehat P(k,k')$, $k,k' \ge 1$ 
constitute a symmetric matrix.

Under conditions that were specified above, configuration 
model networks exhibit a coexistence of a giant component and 
finite tree components.
The set of finite tree components 
constitute a subnetwork which is itself a configuration model network, 
and is in the subcritical regime
\cite{Bollobas2001,Katzav2018}.
Since there are no degree-degree correlations on the finite components,
the joint degree distribution 
of pairs of end-nodes of random edges that reside 
on the finite components
is given by
\begin{equation}
\widehat P(k,k' |{\rm FC}) =
\widetilde P(k|{\rm FC})
\widetilde P(k'|{\rm FC}),
\label{eq:hPkk'FC}
\end{equation}

\noindent
where $k,k' \ge 1$.
Inserting $\widetilde P(k|{\rm FC})$ from Eq. (\ref{eq:tPkFC})
into Eq. (\ref{eq:hPkk'FC}) we obtain
\begin{equation}
\widehat P(k,k' |{\rm FC}) =
(1-\tilde g)^{k+k'-4} \widetilde P(k) \widetilde P(k').
\label{eq:hPkk'FC2}
\end{equation}

\noindent
The joint degree distribution 
$\widehat P(k,k')$
can be expressed 
as a weighted sum of the joint degree distribution of end-nodes
of edges that reside on the giant component and on the finite components
in the form
\begin{equation}
\widehat P(k,k') = 
\widehat P(k,k'|{\rm GC})
\widehat P(e \in {\rm GC})
+
\widehat P(k,k'|{\rm FC})
\widehat P(e \in {\rm FC}),
\label{eq:Pkk'}
\end{equation}

\noindent
where $\widehat P(e \in {\rm GC})$ and $\widehat P(e \in {\rm FC})$
are given by Eqs. (\ref{eq:hPeGC}) and (\ref{eq:hPeFC}), respectively.
Extracting $\widehat P(k,k'|{\rm GC})$ from Eq. (\ref{eq:Pkk'}), we obtain
\begin{equation}
\widehat P(k,k'|{\rm GC}) =
\frac{ \widehat P(k,k') -
\widehat P(k,k'|{\rm FC}) 
\widehat P(e \in {\rm FC})}
{ \widehat P(e \in {\rm GC}) }.
\label{eq:Pkk'GC0}
\end{equation}

\noindent
Note that nodes that reside on the giant component satisfy $k,k' \ge 1$.
Moreover, the giant component does not include edges for which
$k=k'=1$ (dimers), thus $\widehat P(1,1|{\rm GC})=0$.
As a result, the lowest possible degrees of the end-nodes of an edge that resides on the
giant component are $(k,k')=(1,2)$ or $(k,k')=(2,1)$.
This condition can be expressed in the form $k+k' \ge 3$,
in addition to $k,k' \ge 1$.
Inserting the joint degree distributions $\widehat P(k,k')$ and $\widehat P(k,k'|{\rm FC})$
from Eqs. (\ref{eq:hPkk'}) and (\ref{eq:hPkk'FC2}), respectively,
and the probabilities
$\widehat P(e \in {\rm FC})$
and
$\widehat P(e \in {\rm GC})$
from Eqs. (\ref{eq:hPeFC}) and (\ref{eq:hPeGC}),
respectively, into Eq. (\ref{eq:Pkk'GC0}),
we obtain
\begin{equation}
\widehat P(k,k'|{\rm GC}) = 
\frac{  1 -   (1-\tilde g)^{k+k'-2}  }{  (2-\tilde g) \tilde g }
\widetilde P(k) \widetilde P(k'),
\label{eq:Pkk'GC}
\end{equation}

\noindent
where $k,k' \ge 1$ and $k+k' \ge 3$.
Extracting $\widetilde P(k)$ from Eq. (\ref{eq:tPkGC}), we obtain
\begin{equation}
\widetilde P(k) = \frac{ (2-\tilde g)\tilde g }{1 - (1-\tilde g)^k} \widetilde P(k|{\rm GC}).
\label{eq:tPbytPGC}
\end{equation}

\noindent
Inserting $\widetilde P(k)$ and $\widetilde P(k')$ from Eq. (\ref{eq:tPbytPGC})
into Eq. (\ref{eq:Pkk'GC}), we obtain
\begin{eqnarray}
\widehat P(k,k'|{\rm GC}) &=& (2-\tilde g)\tilde g
\frac{ 1 - (1-\tilde g)^{k+k'-2} }{ [1-(1-\tilde g)^k][1-(1-\tilde g)^{k'}] }\nonumber\\
& & \times
\widetilde P(k|{\rm GC}) \widetilde P(k'|{\rm GC}).
\label{eq:hPkk'GCC}
\end{eqnarray}

\noindent
Inserting $k=k'=1$ into Eq. (\ref{eq:hPkk'GCC}), 
we confirm that 
$\widehat P(1,1|{\rm GC})=0$.
In the opposite limit of
$k,k' \rightarrow \infty$,
it is found that
$\widehat P(k,k'|{\rm GC}) \rightarrow (2-\tilde g)\tilde g 
\widetilde P(k|{\rm GC}) \widetilde P(k'|{\rm GC})$.
Since $(2-\tilde g)\tilde g < 1$, the probability
that both end-nodes of an edge that resides on the giant component
will be of high degree is suppressed.
We thus conclude that the degree-degree correlations
between end-nodes of random edges on the giant component
are negative, namely the giant component is disassortative
\cite{Newman2002b,Newman2003a,Newman2003b,Johnson2010,Mizutaka2018}.

\subsection{The covariance of the joint degree distribution of end-nodes of edges}

The 
covariance of the joint degree distribution
of end-nodes of 
edges in a configuration model network
is denoted by
\begin{equation}
\Gamma = 
\widehat{\mathbb{E}}[K K']
-
\widetilde{\mathbb{E}}[K]
\
\widetilde{\mathbb{E}}[K'],
\label{eq:GKK'd}
\end{equation}

\noindent
where
\begin{equation}
\widehat{\mathbb{E}}[K K']
= \sum_{k=1}^{\infty} \sum_{k'=1}^{\infty}
k k' \widehat P(k,k'),
\label{eq:hEKK'd}
\end{equation}

\noindent
is the mixed second moment of $\widehat P(k,k')$.
In configuration model networks there are no degree-degree 
correlations and therefore $\Gamma=0$. 
Moreover, the sub-network that consists of all the
finite tree components is also a configuration model network.
Therefore, the 
covariance of the joint degree distribution of end-nodes of edges
that reside on the 
finite tree components satisfies 
$\Gamma({\rm FC})=0$.

The covariance of the joint degree distribution of
end-nodes of 
edges that reside on the giant component
is given by
\begin{equation}
\Gamma({\rm GC}) = 
\widehat{\mathbb{E}}[KK'|{\rm GC}] - 
\widetilde{\mathbb{E}}[K|{\rm GC}] \  \widetilde{\mathbb{E}}[K'|{\rm GC}],
\label{eq:GKK'GC}
\end{equation}

\noindent
where
\begin{equation}
\widehat{\mathbb{E}}[KK'|{\rm GC}] =
\sum_{k=1}^{\infty} \sum_{k'=1}^{\infty}
k k' \widehat P(k,k'|{\rm GC}) 
\label{eq:hEKK'GC}
\end{equation}

\noindent
is the mixed second moment of
$\widehat P(k,k'|{\rm GC})$
and the mean degree $\widetilde{\mathbb{E}}[K|{\rm GC}]$
is given by Eq. (\ref{eq:tEKGC}).
Inserting $\widehat P(k,k'|{\rm GC})$
from Eq. (\ref{eq:Pkk'GC})
into Eq. (\ref{eq:hEKK'GC}),
carrying out the summations, 
and inserting the result into Eq. (\ref{eq:GKK'GC}),
we obtain
\begin{equation}
\Gamma({\rm GC}) = 
- \frac{ (1-\tilde g)^2 }{  [ (2-\tilde g) \tilde g ]^2 }
\left[ \frac{ \langle K^2 \rangle }{ \langle K \rangle } - 1 - G'_1(1-\tilde g)  \right]^2.
\label{eq:GGC}
\end{equation}

\noindent
It is found that $\Gamma({\rm GC}) < 0$, namely the giant component of 
a configuration model network is always disassortative
\cite{Newman2002b,Newman2003a,Newman2003b,Johnson2010,Mizutaka2018}.
This means that on the giant component high degree nodes tend to connect
to low degree nodes and vice versa.

In the dilute network regime of $0 < \tilde g \ll 1$, 
just above the percolation transition,
the giant component is small
but it exhibits strong degree-degree correlations.
Using Eq. (\ref{eq:G1'b}), it is found that in this regime
\begin{equation}
\Gamma({\rm GC}) \simeq 
- \frac{1}{4}
\left( \frac{ \langle K^3 \rangle }{ \langle K \rangle }
- 3 \frac{ \langle K^2 \rangle }{ \langle K \rangle } + 2 \right)^2
+ \mathcal{O}\left( \tilde g^2 \right).
\label{eq:GGCgll1}
\end{equation}

\noindent
In the opposite limit of $\tilde g \rightarrow 1^{-}$, in which the giant component expands
to encompass the whole network
(apart from any isolated nodes), 
$\Gamma({\rm GC}) \rightarrow 0$.
More precisely, in the regime of $1 - \tilde g \ll 1$
the covariance of the joint degree distribution of end-nodes on the giant component
decays according to
$\Gamma({\rm GC}) \sim  - (1-\tilde g)^2$.
The Pearson correlation coefficient for pairs of end-nodes of edges that reside
on the giant component is given by
\begin{equation}
R({\rm GC}) = \frac{\Gamma({\rm GC})}{\widetilde{\mathbb{V}}[K|{\rm GC}]},
\label{eq:RGC}
\end{equation}

\noindent
where $\widetilde{\mathbb{V}}[K|{\rm GC}]$
is given by Eq. (\ref{eq:tVKGC}).
Unlike the covariance $\Gamma({\rm GC})$, the
Pearson correlation coefficient is bounded in the range
$-1 \le R({\rm GC}) \le 1$.
It is thus a more suitable measure for the comparison of the
correlations between the degrees of pairs of end-nodes in different populations of edges and bredges.

\begin{widetext}

\section{Statistical analysis of bredges}

\subsection{The probability that a random edge is a bredge}

Consider a random edge $e$ in a configuration model network of $N$
nodes with degree distribution $P(k)$. 
The probability $\widehat P(e \in {\rm B})$ that $e$ is a bredge is given by
\cite{Wu2018}
\begin{equation}
\widehat P(e \in {\rm B}) = 1 - \tilde g^2,
\label{eq:hPeB}
\end{equation}

\noindent
where $\tilde g$ is given by Eq. (\ref{eq:tg}). 
This is due to the fact that in order that a random edge will not be
a bredge, its end-nodes $i$ and $i'$ should both belong
to the giant component of the reduced network from which the edge $e$ 
was removed. The probability for each one of these nodes to belong to the
giant component of the reduced network is $\tilde g$.
Thus, the probability that both of them belong to the giant component of the reduced network
is $\tilde g^2$. The probability that at least one of them does not belong to the giant component
of the reduced network is thus $1-\tilde g^2$, which leads to Eq. (\ref{eq:hPeB}).
The complementary probability, that a random edge $e$ is a non-bredge (NB) edge
is given by
$\widehat P(e \in {\rm NB})=\tilde g^2$.
Therefore, in the dilute network regime of $0 < \tilde g \ll 1$, just above the percolation transition,
almost every edge is a bredge.

Consider a random edge $e$ whose end-nodes $i$ and $i'$  
are of known degrees, $k$ and $k'$, where $k,k' \ge 1$. 
In order that the edge $e$ will not be a bredge, both $i$ and $i'$
should reside on the giant component of the reduced network from
which $e$ is removed.
Therefore, the probability that $e$ is
a bredge is given by
\begin{equation}
\widehat P(e \in {\rm B}| k, k') =
1 - [1-(1-\tilde g)^{k-1}][1-(1-\tilde g)^{k'-1}],
\label{eq:hPeBkk'0}
\end{equation}

\noindent
where $k,k' \ge 1$.
This result can also be expressed in the form
\begin{equation}
\widehat P(e \in {\rm B}| k, k') =
(1-\tilde g)^{k-1} + (1-\tilde g)^{k'-1} 
- (1-\tilde g)^{k+k'-2}.
\label{eq:hPeBkk'}
\end{equation}

\noindent
The probability 
$\widehat P(e \in {\rm B})$
that a random edge is a bredge can be expressed in the form
\begin{equation}
\widehat P(e \in {\rm B}) =
\sum_{k,k'=1}^{\infty}
\widehat P(e \in {\rm B}| k,k') \widehat P(k,k').
\label{eq:hPeBd}
\end{equation}

\noindent
Inserting the conditional probability $\widehat P(e \in {\rm B}|k,k')$ from
Eq. (\ref{eq:hPeBkk'}) into Eq. (\ref{eq:hPeBd}) and carrying out the summation,
one recovers Eq. (\ref{eq:hPeB}).

The probability $\widehat P(e \in {\rm B})$ can be expressed as a sum of two terms,
where one term accounts for nodes that reside on the giant component and the other
accounts for nodes that reside on the finite components.
It takes the form
\begin{equation}
\widehat P(e \in {\rm B}) = 
\widehat P(e \in {\rm B}|{\rm GC}) \widehat P(e \in {\rm GC})
+
\widehat P(e \in {\rm B}|{\rm FC}) \widehat P(e \in {\rm FC}).
\label{eq:hPeBd2}
\end{equation}

\noindent
Extracting the conditional probability
$\widehat P(e \in {\rm B}|{\rm GC})$ 
that a random edge on the giant component is a bredge,
one obtains
\begin{equation}
\widehat P(e \in {\rm B}|{\rm GC}) 
=
\frac{ \widehat P(e \in {\rm B}) - \widehat P(e \in {\rm B}|{\rm FC}) \widehat P(e \in {\rm FC}) }
{ \widehat P(e \in {\rm GC}) }.
\label{eq:hPeBGCd}
\end{equation}

\noindent
Since all the edges on the finite tree components are bredges,
$\widehat P(e \in {\rm B}|{\rm FC})=1$.
Inserting 
$\widehat P(e \in {\rm B})$
from Eq. (\ref{eq:hPeB}),
$\widehat P(e \in {\rm GC})$
from Eq. (\ref{eq:hPeGC})
and
$\widehat P(e \in {\rm FC})$
from Eq. (\ref{eq:hPeFC})
into Eq. (\ref{eq:hPeBGCd}), 
we obtain
\begin{equation}
\widehat P(e \in {\rm B}|{\rm GC}) 
=
\frac{ 2(1-\tilde g) }{2 - \tilde g}.
\label{eq:hPeBGC}
\end{equation}

\noindent
Therefore, the complementary probability
that a random edge on the giant component 
is not a bredge is given by
\begin{equation}
\widehat P(e \in {\rm NB}|{\rm GC}) 
=
\frac{ \tilde g }{2 - \tilde g}.
\label{eq:hPeNBGC}
\end{equation}

To calculate the fraction of bredges that belong to the giant component
one can use Bayes' theorem, and obtain
\begin{equation}
\widehat P(e \in {\rm GC}|{\rm B}) = 
\frac{ \widehat P(e \in {\rm B}|{\rm GC}) \widehat P(e \in {\rm GC}) }
{ \widehat P(e \in {\rm B}) }.
\label{eq:hPeGCB0}
\end{equation}

\noindent
Inserting 
$\widehat P(e \in {\rm B}|{\rm GC})$
from Eq. (\ref{eq:hPeBGC}),
$\widehat P(e \in {\rm GC})$
from Eq. (\ref{eq:hPeGC})
and
$\widehat P(e \in {\rm B})$
from Eq. (\ref{eq:hPeB})
into Eq. (\ref{eq:hPeGCB0}),
we obtain
\begin{equation}
\widehat P(e \in {\rm GC}|{\rm B}) = 
\frac{ 2 \tilde g }{1+\tilde g}.
\label{eq:hPeGCB}
\end{equation}

\noindent
Therefore, the fraction of bredges that reside on the finite components is
\begin{equation}
\widehat P(e \in {\rm FC}|{\rm B}) 
=
\frac{ 1 - \tilde g }{1 + \tilde g}.
\label{eq:hPeFCB}
\end{equation}

The conditional probability $P(e \in {\rm B}|k,k')$,
given by Eq. (\ref{eq:hPeBkk'}),
can be expressed as a sum of two terms,
where one term accounts for nodes that reside on the giant component and the other
accounts for nodes that reside on the finite components.
It takes the form
\begin{eqnarray}
\widehat P(e \in {\rm B}|k,k') &=& 
\widehat P(e \in {\rm B}|{\rm GC},k,k') \widehat P(e \in {\rm GC}|k,k')
\nonumber \\
&+&
\widehat P(e \in {\rm B}|{\rm FC},k,k') \widehat P(e \in {\rm FC}|k,k').
\label{eq:hPeBkk'2}
\end{eqnarray}

\noindent
The conditional probability $\widehat P(e \in {\rm GC}|k,k')$,
given by Eq. (\ref{eq:hPeGCkk'}),
takes non-zero values only for degrees
$k,k' \ge 1$ whose sum satisfies
$k+k' \ge 3$,
while 
$\widehat P(e \in {\rm FC}|k,k')$
is given by Eq. (\ref{eq:hPeFCkk'}),
where $k,k' \ge 1$.
Since all the edges that reside on the finite components are bredges,
$\widehat P(e \in {\rm B}|{\rm FC},k,k')=1$.
Extracting the conditional probability
$\widehat P(e \in {\rm B}|{\rm GC},k,k')$ 
from Eq. (\ref{eq:hPeBkk'2}),
one obtains
\begin{equation}
\widehat P(e \in {\rm B}|{\rm GC},k,k') 
=
\frac{ \widehat P(e \in {\rm B}|k,k') - \widehat P(e \in {\rm B}|{\rm FC},k,k') \widehat P(e \in {\rm FC}|k,k') }
{ \widehat P(e \in {\rm GC}|k,k') }.
\label{eq:hPeBGCkk'}
\end{equation}

\noindent
Since all the edges on the finite tree components are bredges,
$\widehat P(e \in {\rm B}|{\rm FC},k,k')=1$, where $k,k' \ge 1$.
Evaluating the right hand side of Eq. (\ref{eq:hPeBGCkk'}), we obtain
\begin{equation}
\widehat P(e \in {\rm B}|{\rm GC},k,k') 
=
\frac{ (1-\tilde g)^{k-1} 
+
(1-\tilde g)^{k'-1}
-
2 (1-\tilde g)^{k+k'-2} }
{ 1 - (1-\tilde g)^{k+k'-2} },
\label{eq:hPeBGCkk'2}
\end{equation}

\noindent
where $k,k' \ge 1$ and $k+k' \ge 3$.
The probability that an edge connecting end-nodes of degrees $k$ and $k'$ on the giant component
is not a bredge is thus given by
\begin{equation}
\widehat P(e \in {\rm NB}|{\rm GC},k,k') 
=
\frac{ 1 - (1-\tilde g)^{k-1} 
-
(1-\tilde g)^{k'-1}
+
(1-\tilde g)^{k+k'-2} }
{ 1 - (1-\tilde g)^{k+k'-2} },
\label{eq:hPeNBGCkk'2}
\end{equation}

\noindent
where $k,k' \ge 1$ and $k+k' \ge 3$.

\subsection{The joint degree distribution of the end-nodes of bredges}

The joint degree distribution of the nodes on both sides
of a bredge can be expressed in the form
\begin{equation}
\widehat P(k,k'|{\rm B}) =
\frac{ \widehat P(e \in {\rm B}|k,k')  \widehat P(k,k') }{\widehat P(e \in {\rm B})}.
\label{eq:hPkk'B0}
\end{equation}

\noindent
Inserting $\widehat P(e \in {\rm B}|k,k')$
from Eq. (\ref{eq:hPeBkk'}),
$\widehat P(k,k')$
from Eq. (\ref{eq:hPkk'})
and $\widehat P(e \in {\rm B})$
from Eq. (\ref{eq:hPeB})
into Eq. (\ref{eq:hPkk'B0}), 
we obtain
\begin{equation}
\widehat P(k,k'|{\rm B}) =
\frac{1}{1+\tilde g}
\left[ 
(1-\tilde g)^{k-2}
+
(1-\tilde g)^{k'-2}
-
(1-\tilde g)^{k+k'-3}
\right]
\widetilde P(k) \widetilde P(k'),
\label{eq:hPkk'B}
\end{equation}

\noindent
where $k,k' \ge 1$.
Below we consider the joint degree distributions 
$\widehat P(k,k'|{\rm B},{\rm GC})$ and 
$\widehat P(k,k'|{\rm B},{\rm FC})$
of the end-nodes
of random bredges on the giant component
and on the finite components, respectively.
Since all the edges on the finite components are bredges,
the joint degree distribution of the end nodes of random bredges
that reside on the finite components satisfies
$\widehat P(k,k'|{\rm B},{\rm FC}) = \widehat P(k,k'|{\rm FC})$,
where $\widehat P(k,k'|{\rm FC})$
is given by Eq. (\ref{eq:hPkk'FC2}).

The conditional probability $\widehat P(k,k'|{\rm B})$ can be expressed in the form
\begin{eqnarray}
\widehat P(k,k' |{\rm B}) &=& 
\widehat P(k,k'|{\rm B},{\rm GC}) \widehat P(e \in {\rm GC}|{\rm B})
\nonumber \\
&+&
\widehat P(k,k'|{\rm B},{\rm FC}) \widehat P(e \in {\rm FC}|{\rm B}).
\label{eq:hPkk'B2}
\end{eqnarray}

\noindent
Extracting $\widehat P(k,k'|{\rm B},{\rm GC})$ 
from Eq. (\ref{eq:hPkk'B2})
we obtain
\begin{equation}
\widehat P(k,k'|{\rm B},{\rm GC}) =
\frac{ \widehat P(k,k' |{\rm B})
-
\widehat P(k,k'|{\rm B},{\rm FC}) \widehat P(e \in {\rm FC}|{\rm B}) }
{ \widehat P(e \in {\rm GC}|{\rm B}) }.
\label{eq:hPkk'BGC0}
\end{equation}

\noindent
Inserting
$\widehat P(e \in {\rm FC}|{\rm B})$
from Eq. (\ref{eq:hPeFCB})
and
$\widehat P(e \in {\rm GC}|{\rm B})$
from Eq. (\ref{eq:hPeBGC})
into Eq. (\ref{eq:hPkk'BGC}),
we obtain
\begin{equation}
\widehat P(k,k'|{\rm B},{\rm GC}) =
\frac{1}{ 2 \tilde g }
\left[ (1-\tilde g)^{k-2} + (1-\tilde g)^{k'-2} - 2(1-\tilde g)^{k+k'-3} \right]
\widetilde P(k) \widetilde P(k'),
\label{eq:hPkk'BGC}
\end{equation}

\noindent
where $k,k' \ge 1$ and $k+k' \ge 3$.

The joint degree distribution of pairs of end-nodes 
of random edges on the giant component
can be decomposed in the form
\begin{equation}
\widehat P(k,k'|{\rm GC}) =
\widehat P(k,k'|{\rm B},{\rm GC}) \widehat P(e \in {\rm B}|{\rm GC})
+
\widehat P(k,k'|{\rm NB},{\rm GC}) \widehat P(e \in {\rm NB}|{\rm GC}),
\end{equation}

\noindent
where the first term on the right hand side accounts for the bredges and
the second term account for all the edges that are not bredges.
Note that the joint degree distribution
$\widehat P(k,k'|{\rm NB},{\rm GC})$
may take non-zero values only under conditions in
which both $k \ge 2$ and $k' \ge 2$.
Extracting the joint degree distribution on the edges that are not bredges,
we obtain
\begin{equation}
\widehat P(k,k'|{\rm NB},{\rm GC})
=
\frac{
\widehat P(k,k'|{\rm GC})  - 
\widehat P(k,k'|{\rm B},{\rm GC}) \widehat P(e \in {\rm B}|{\rm GC}) }
{ \widehat P(e \in {\rm NB}|{\rm GC}) }.
\end{equation}

\noindent
Inserting
$\widehat P(k,k'|{\rm GC})$
from Eq. (\ref{eq:Pkk'GC}),
$\widehat P(k,k'|{\rm B},{\rm GC})$
from Eq. (\ref{eq:hPkk'BGC}),
$\widehat P(e \in {\rm B}|{\rm GC})$
from Eq. (\ref{eq:hPeBGC})
and 
$\widehat P(e \in {\rm NB}|{\rm GC})$
from Eq. (\ref{eq:hPeNBGC}), 
we obtain
\begin{equation}
\widehat P(k,k'|{\rm NB},{\rm GC})
=
\frac{1}{\tilde g^2}
\left[ 1 - (1-\tilde g)^{k-1} - (1-\tilde g)^{k'-1} + (1-\tilde g)^{k+k'-2} \right]
\widetilde P(k) \widetilde P(k'),
\label{eq:hPkk'NBGC}
\end{equation}

\noindent
where $k,k' \ge 2$.
Eq. (\ref{eq:hPkk'NBGC}) can be written as a product of the form
\begin{equation}
\widehat P(k,k'|{\rm NB},{\rm GC})
=
\left[ \frac{   1 - (1-\tilde g)^{k-1}   }{\tilde g} \right]
\widetilde P(k) 
\left[ \frac{  1 - (1-\tilde g)^{k'-1}  }{\tilde g} \right]
\widetilde P(k'),
\label{eq:hPkk'NBGC2}
\end{equation}

\noindent
which means that the degrees $k$ and $k'$ of the end-nodes of non-bredge edges
on the giant component are uncorrelated.
Therefore, the degree distribution 
$\widetilde P(k|{\rm NB},{\rm GC})$
of end-nodes of non-bredge edges that reside
on the giant component 
is given by
\begin{equation}
\widetilde P(k|{\rm NB},{\rm GC}) = 
\frac{ 1 -  (1-\tilde g)^{k-1} }{ \tilde g}
\widetilde P(k),
\label{eq:tPkNBGC}
\end{equation}

\noindent
where $k \ge 2$.
We thus conclude that all the degree-degree correlations in 
the giant component of a
configuration model
network are concentrated in the bredges.

A special property of bredges on the giant component is that 
they are `polarized' in the sense that
each bredge $e$
has one end-node that resides on the giant component of the reduced network
from which $e$ is removed, while the other end-node resides on the finite tree
component that is detached from the giant component.
These two end-nodes exhibit different statistical properties.
The conditional probability that the end-node $i$ (of degree $k$) resides on the
giant component and the end-node $i'$ (of degree $k'$) resides on the detached finite tree 
is given by
\begin{equation}
\widehat P(K_{\rm GC}=k,K_{\rm FC}=k'|k,k',{\rm B},{\rm GC})
=
\frac{ (1-\tilde g)^{k'-1} \left[ 1 - (1-\tilde g)^{k-1} \right] }
{ (1-\tilde g)^{k-1} + (1-\tilde g)^{k'-1} - 2(1-\tilde g)^{k+k'-2} },
\label{eq:hPkGkFc}
\end{equation}

\noindent
for $k \ne k'$ 
and
$\widehat P(K_{\rm GC}=k,K_{\rm FC}=k'|k,k',{\rm B},{\rm GC})=1$
for $k=k'$.
The joint degree distribution
$\widehat P(K_{\rm GC}=k,K_{\rm FC}=k'|{\rm B},{\rm GC})$
can thus be written in the form
\begin{equation}
\widehat P(K_{\rm GC}=k,K_{\rm FC}=k'|{\rm B},{\rm GC}) =
(2-\delta_{k,k'})
\widehat P(K_{\rm GC}=k,K_{\rm FC}=k'|k,k',{\rm B},{\rm GC})
\widehat P(k,k'|{\rm B},{\rm GC}),
\label{eq:hPkGkF}
\end{equation}

\noindent
where $\delta_{k,k'}$ is the Kronecker delta symbol
and $\widehat P(k,k'|{\rm B},{\rm GC})$ 
is given by Eq. (\ref{eq:hPkk'BGC}).
Inserting the right hand side of Eq. (\ref{eq:hPkGkFc}) into Eq. (\ref{eq:hPkGkF}),
we find that
Eq. (\ref{eq:hPkGkF})
can be written 
as a product of the form
\begin{equation}
\widehat P(K_{\rm GC}=k,K_{\rm FC}=k'|{\rm B},{\rm GC}) =
\widetilde P(K_{\rm GC}=k|{\rm B},{\rm GC})  
\widetilde P(K_{\rm FC}=k'|{\rm B},{\rm GC}),
\label{eq:hPKGkKFk'd}
\end{equation}

\noindent
where the the degree distribution of the end-node on the giant component side is
\begin{equation}
\widetilde P(K_{\rm GC}=k|{\rm B},{\rm GC}) =
\frac{ 1 - (1-\tilde g)^{ k-1}  }{\tilde g}
\widetilde P(k),
\label{eq:tPKGkBGC}
\end{equation}

\noindent
and the degree distribution of the end-node on the finite component side is
\begin{equation}
\widetilde P(K_{\rm FC}=k'|{\rm B},{\rm GC}) =
(1-\tilde g)^{k'-2} \widetilde P(k').
\label{eq:tPKFk'BGC}
\end{equation}

\noindent
Eq. (\ref{eq:hPKGkKFk'd})
implies that once we recognize that each bredge
on the giant component
has one end-node whose degree is sampled from
$\widetilde P(K_{\rm GC}=k|{\rm B},{\rm GC})$,
while the degree of the other end-node is sampled from
$\widetilde P(K_{\rm FC}=k'|{\rm B},{\rm GC})$,
the correlation between the degrees of the two end-nodes vanishes.
The correlation found in the analysis above, 
between the degrees $k$ and $k'$ 
of the end-nodes $i$ and $i'$,
in the joint degree distribution $\widehat P(k,k'|{\rm B},{\rm GC})$ 
[Eq. (\ref{eq:hPkk'BGC})] is due to the fact that if $i$ 
ends up on the giant component of the reduced network,
then $i'$ must end up on a finite component and vice versa.

\subsection{The marginal degree distribution of the end-nodes of bredges}

The degree distribution 
$\widetilde P(k|B)$
of an end-node of a random bredge can be 
obtained as the marginal distribution of 
the joint degree distribution
$\widehat P(k,k'|B)$ by tracing over $k'$, namely
\begin{equation}
\widetilde P(k|{\rm B}) = \sum_{k'=1}^{\infty} \widehat P(k,k'|{\rm B}).
\label{eq:tPkBd}
\end{equation}

\noindent
Inserting $\widehat P(k,k'|{\rm B})$ from Eq. (\ref{eq:hPkk'B}) 
and carrying out the summation,
we obtain
\begin{equation}
\widetilde P(k|{\rm B}) =
\frac{ 1 + \tilde g (1-\tilde g)^{k-2} }{1+\tilde g}
\widetilde P(k).
\label{eq:tPkB}
\end{equation}

\noindent
Extracting $\widetilde P(k)$ from Eq. (\ref{eq:tPkB}),
we obtain
\begin{equation}
\widetilde P(k) =
\frac{1+\tilde g}{ 1 + \tilde g (1-\tilde g)^{k-2} }
\widetilde P(k|{\rm B}).
\label{eq:tPkB2}
\end{equation}

\noindent
Inserting $\widetilde P(k)$ and $\widetilde P(k')$
from Eq. (\ref{eq:tPkB2}) into Eq. (\ref{eq:hPkk'B}), we obtain
\begin{equation}
\widehat P(k,k'|{\rm B}) =
(1+\tilde g)
\frac{ (1-\tilde g)^{k-2} + (1-\tilde g)^{k'-2} - (1-\tilde g)^{k+k'-3} }
{ \left[ 1 + \tilde g(1-\tilde g)^{k-2} \right] \left[ 1 + \tilde g(1-\tilde g)^{k'-2} \right] }
\widetilde P(k|{\rm B}) \widetilde P(k'|{\rm B}).
\end{equation}

Since on the finite tree components all the edges are bredges, 
the degree distribution $\widetilde P(k|{\rm B},{\rm FC})$ of the end-nodes
of bredges that reside on the finite components satisfies
$\widetilde P(k|{\rm B},{\rm FC}) = \widetilde P(k|{\rm FC})$,
where $\widetilde P(k|{\rm FC})$ is given by Eq. (\ref{eq:tPkFC}).
The degree distribution 
$\widetilde P(k|{\rm B},{\rm GC})$
of end-nodes of bredges that reside
on the giant component can be 
obtained by marginalizing 
$\widehat P(k,k'|{\rm B},{\rm GC})$,
given by Eq. (\ref{eq:hPkk'BGC}),
over $k'$.
This yields
\begin{equation}
\widetilde P(k|{\rm B},{\rm GC}) = 
\frac{ 1 + (2 \tilde g-1)(1-\tilde g)^{k-2} }{2 \tilde g}
\widetilde P(k),
\label{eq:tPkBGC}
\end{equation}

\noindent
where $k \ge 1$.
Extracting $\widetilde P(k)$ from Eq. (\ref{eq:tPkBGC}),
we obtain
\begin{equation}
\widetilde P(k) =
\frac{2 \tilde g}{ 1 + (2 \tilde g-1)(1-\tilde g)^{k-2} }
\widetilde P(k|{\rm B},{\rm GC}). 
\label{eq:tPkBGC2}
\end{equation}

\noindent
Inserting $\widetilde P(k)$ from Eq. (\ref{eq:tPkBGC2})
into Eq. (\ref{eq:hPkk'BGC}), we obtain
\begin{equation}
\widehat P(k,k'|{\rm B},{\rm GC}) =
\frac{ 2 \tilde g \left[ (1-\tilde g)^{k-2} + (1-\tilde g)^{k'-2} - 2(1-\tilde g)^{k+k'-3} \right]   }
{  \left[ 1 + (2 \tilde g-1)(1-\tilde g)^{k-2}  \right] \left[ 1 + (2 \tilde g-1)(1-\tilde g)^{k'-2} \right] }
\widetilde P(k|{\rm B},{\rm GC}) \widetilde P(k'|{\rm B},{\rm GC}),
\label{eq:hPkk'BGC2}
\end{equation}

\subsection{The mean degree of end-nodes of bredges}

The mean degree of end-nodes of bredges is denoted by
$\widetilde{ \mathbb{E}}[K|{\rm B}]$.
Using the degree distribution $\widetilde P(k|{\rm B})$, given by Eq. (\ref{eq:tPkB}),
we obtain
\begin{equation}
\widetilde{ \mathbb{E}}[K|{\rm B}]  
= 
\frac{1}{1+\tilde g}
\left\{ \frac{ \langle K^2 \rangle }{\langle K \rangle}
+
\tilde g \big[ 1 + G'_1(1-\tilde g) \big]
\right\}.
\label{eq:tEKB}
\end{equation}

\noindent
The mean degree 
$\widetilde{ \mathbb{E}}[K|{\rm B},{\rm FC}]$
of the end-nodes of random bredges that reside on  
the finite components is
identical to
$\widetilde{ \mathbb{E}}[K|{\rm FC}]$,
which is given by Eq. (\ref{eq:tEKFC}).
Using Eq. (\ref{eq:G1'}) it is found that
$\widetilde{ \mathbb{E}}[K|{\rm B},{\rm FC}] < \widetilde{ \mathbb{E}}[K]$.
Using Eq. (\ref{eq:tPkBGC}) we obtain the mean degree of the end-nodes of bredges
that reside on the giant component, which is given by
\begin{equation}
\widetilde{ \mathbb{E}}[K|{\rm B},{\rm GC}] =
\frac{1}{2 \tilde g}
\left\{
\frac{ \langle K^2 \rangle }{ \langle K \rangle }
+
(2 \tilde g - 1) \big[ 1 + G_1'(1-\tilde g) \big]
\right\}.
\label{eq:tEKBGC}
\end{equation}

\noindent
Using Eq. (\ref{eq:G1'}) it is found that
$\widetilde{ \mathbb{E}}[K|{\rm B},{\rm GC}] \ge \widetilde{ \mathbb{E}}[K]$.
Note that in heavy tail degree distributions
the mean degree 
$\widetilde{ \mathbb{E}}[K|{\rm B},{\rm GC}]$ 
of end-nodes on the giant component
may diverge even under conditions in which
$\langle K \rangle$ is finite.
This is due to the fact that the second moment
$\langle K^2 \rangle$ appears on the right hand side
of Eq. (\ref{eq:tEKGC}).
In heavy-tail degree distributions $\langle K^2 \rangle$
may diverge, leading to the divergence of
$\widetilde{\mathbb{E}}[K|{\rm B},{\rm GC}]$.

Below we evaluate the means
of the degree distributions 
of the end-nodes of bredges $e$ on the giant component,
which reside on the giant and on the finite components 
of the reduced network from which $e$ is removed.
Using Eq. (\ref{eq:tPKGkBGC})
we obtain the mean degree of the end-nodes that reside
on the giant component of the reduced network,
which is given by
\begin{equation}
\widetilde{\mathbb{E}}[K_{\rm GC}|{\rm B},{\rm GC}] 
=
\frac{1}{\tilde g}
\left\{ \frac{ \langle K^2 \rangle }{\langle K \rangle}
-
(1-\tilde g)  \big[ 1 + G'_1(1-\tilde g) \big] \right\}.
\label{eq:tEKGBGC}
\end{equation}

\noindent
Using
Eq. (\ref{eq:tPKFk'BGC})
we obtain the mean degree of the end-nodes that reside on a
finite component of the reduced network,
which is given by
\begin{equation}
\widetilde{\mathbb{E}}[K_{\rm FC}|{\rm B},{\rm GC}] 
=
1 + G'_1(1-\tilde g).
\label{eq:tEKFBGC2} 
\end{equation}

\noindent
Similarly, the mean degree of the end-nodes of random non-bredge edges that reside on  
the giant component,
obtained using Eq. (\ref{eq:tPkNBGC}),
is given by
\begin{equation}
\widetilde{\mathbb{E}}[K|{\rm NB},{\rm GC}]
=
\frac{1}{\tilde g}
\left\{ \frac{ \langle K^2 \rangle }{ \langle K \rangle }
- (1-\tilde g)  \big[1 + G'_1(1-\tilde g) \big]
\right\}.
\label{eq:tEKNBGC}
\end{equation}

\subsection{The variance of the degree distribution of end-nodes of bredges}

The second moment of the degree distribution
$\widetilde P(k|{\rm B})$ of the end-nodes of bredges,
obtained using Eq. (\ref{eq:tPkB}), 
is given by
\begin{equation}
\widetilde{\mathbb{E}}[K^2|{\rm B}]
= 
\frac{1}{1+\tilde g}
\left\{ \frac{ \langle K^3 \rangle }{\langle K \rangle}  
+
\tilde g \big[ (1-\tilde g) G''_1(1-\tilde g) + 3 G'_1(1-\tilde g) + 1 \big] \right\}.
\label{eq:tEK2B}
\end{equation}

\noindent
Using
$\widetilde{ \mathbb{E}}[K|{\rm B}]$
from Eq. (\ref{eq:tEKB})
and
$\widetilde{ \mathbb{E}}[K^2|{\rm B}]$
from Eq. (\ref{eq:tEK2B}),
we obtain the variance
\begin{eqnarray}
\widetilde{\mathbb V}[K|{\rm B}] &=& 
\frac{1}{1+\tilde g}
\left\{ \frac{ \langle K^3 \rangle }{\langle K \rangle} 
+
\tilde g \big[ (1-\tilde g) G''_1(1-\tilde g) + 3 G'_1(1-\tilde g) + 1 \big] \right\}
\nonumber \\
&-&
\frac{1}{(1+\tilde g)^2}
\left\{ \frac{ \langle K^2 \rangle }{\langle K \rangle}
+
\tilde g \big[ 1 + G'_1(1-\tilde g) \big]
\right\}^2.
\label{eq:tVKB}
\end{eqnarray}

\noindent
Since all the edges on the finite components are bredges, it is clear that
$\widetilde{\mathbb{E}}[K^2|{\rm B},{\rm FC}]=\widetilde{\mathbb{E}}[K^2|{\rm FC}]$,
which is given by Eq. (\ref{eq:tEK2FC}).
Similarly, 
$\widetilde{\mathbb{V}}[K|{\rm B},{\rm FC}] = \widetilde{\mathbb{V}}[K|{\rm FC}]$,
which is given by Eq. (\ref{eq:tVKFC}).
The second moment of the degree distribution
$\widetilde P(k|{\rm B},{\rm GC})$
of nodes selected via random edges that reside on  
the giant component,
obtained using Eq. (\ref{eq:tPkB}),
is given by
\begin{equation}
\widetilde{\mathbb{E}}[K^2|{\rm B},{\rm GC}]
=
\frac{1}{2 \tilde g}
\left\{
\frac{\langle K^3 \rangle}{\langle K \rangle}
+ (2 \tilde g - 1)
\big[ (1-\tilde g) G''_1(1-\tilde g) + 3 G'_1(1-\tilde g) + 1 \big]
\right\}.
\label{eq:tEK2BGC}
\end{equation}

\noindent
Using the first and second moments from Eqs. (\ref{eq:tEKBGC}) and (\ref{eq:tEK2BGC}),
respectively, we obtain the variance
\begin{eqnarray}
\widetilde{\mathbb{V}}[K|{{\rm B},\rm GC}] 
&=&  
\frac{1}{2 \tilde g}
\left\{
\frac{\langle K^3 \rangle}{\langle K \rangle}
+ (2 \tilde g - 1)
\big[ (1-\tilde g) G''_1(1-\tilde g) + 3 G'_1(1-\tilde g) + 1 \big]
\right\}
\nonumber \\
&-&
\frac{1}{4 \tilde g^2}
\left\{
\frac{ \langle K^2 \rangle }{ \langle K \rangle }
+
(2 \tilde g - 1) \big[ 1 + G_1'(1-\tilde g) \big]
\right\}^2
\label{eq:tVKBGC}
\end{eqnarray}

Below we evaluate the second moments
of the degree distributions 
and
of the end-nodes of bredges $e$ on the giant component,
which end up on the giant and on the finite components 
of the reduced network from which $e$ is removed.
The second moment of the degree distribution of the end-nodes that end up on the
giant component,
obtained using Eq. (\ref{eq:tPKGkBGC}),
is given by
\begin{equation}
\widetilde{\mathbb{E}}[K_{\rm GC}^2|{\rm B},{\rm GC}] 
=
\frac{1}{\tilde g}
\left\{ \frac{ \langle K^3 \rangle }{\langle K \rangle}
-
(1-\tilde g)^2 \big[ G''_1(1-\tilde g) + 3 G'_1(1-\tilde g) + 1 \big] \right\}.
\label{eq:tEKG2BGC} 
\end{equation}

\noindent
Using the first and second moments from Eqs. (\ref{eq:tEKGBGC}) and (\ref{eq:tEKG2BGC}),
respectively, we obtain the variance
\begin{eqnarray}
\widetilde{\mathbb{V}}[K_{\rm GC}|{{\rm B},\rm GC}] 
&=&  
\frac{1}{\tilde g}
\left\{ \frac{ \langle K^3 \rangle }{\langle K \rangle}
-
(1-\tilde g)^2 \big[ G''_1(1-\tilde g) + 3 G'_1(1-\tilde g)+1 \big] \right\}
\nonumber \\
&-&
\frac{1}{\tilde g^2}
\left\{ \frac{ \langle K^2 \rangle }{\langle K \rangle}
-
(1-\tilde g) \big[ 1 + G'_1(1-\tilde g) \big] \right\}^2.
\label{eq:tVKGCBGC}
\end{eqnarray}

\noindent
The second moment of the degree distribution of the end-nodes that end up on the
finite tree that is detached from the giant component of the reduced network,
obtained using Eq. (\ref{eq:tPKFk'BGC}),
is given by
\begin{equation}
\widetilde{\mathbb{E}}[K_{\rm FC}^2|{\rm B},{\rm GC}] 
=
(1-\tilde g) G''_1(1-\tilde g) + 3 G'_1(1-\tilde g) + 1.
\label{eq:tEKF2BGC2} 
\end{equation}

\noindent
Using the first and second moments from Eqs. (\ref{eq:tEKFBGC2}) and (\ref{eq:tEKF2BGC2}),
respectively, we obtain the variance

\begin{eqnarray}
\widetilde{\mathbb{V}}[K_{\rm FC}|{{\rm B},\rm GC}] 
&=&  
(1-\tilde g) G''_1(1-\tilde g) 
+ G'_1(1-\tilde g) \big[ 1 - G'_1(1-\tilde g) \big].
\label{eq:tVKBGC2}
\end{eqnarray}

\noindent
The second moment of the degree distribution
$\widetilde P(k|{\rm NB},{\rm GC})$ of the end-nodes of non-bredge edges 
that reside on the giant component,
obtained using Eq. (\ref{eq:tPkNBGC}),
is given by
\begin{equation}
\widetilde{\mathbb{E}}[K^2|{\rm NB},{\rm GC}]
= 
\frac{1}{\tilde g}
\left\{ \frac{ \langle K^3 \rangle }{\langle K \rangle} 
-
(1-\tilde g)
\big[ (1-\tilde g) G''_1(1-\tilde g) + 3 G'_1(1-\tilde g) + 1 \big]
\right\}.
\label{eq:tEK2NBGC}
\end{equation}

\noindent
using
$\widetilde{ \mathbb{E}}[K|{\rm NB},{\rm GC}]$
from Eq. (\ref{eq:tEKNBGC})
and
$\widetilde{ \mathbb{E}}[K^2|{\rm NB},{\rm GC}]$
from Eq. (\ref{eq:tEK2NBGC}),
we obtain the variance
\begin{eqnarray}
\widetilde {\mathbb V}[K|{\rm NB},{\rm GC}] &=& 
\frac{1}{\tilde g}
\left\{ \frac{ \langle K^3 \rangle }{\langle K \rangle} 
-
(1-\tilde g)
\big[ (1-\tilde g) G''_1(1-\tilde g) + 3 G'_1(1-\tilde g) + 1 \big]
\right\}
\nonumber \\
&-&
\frac{1}{\tilde g^2}
\left\{ \frac{ \langle K^2 \rangle }{\langle K \rangle}
-
(1-\tilde g) \big[ 1 + G'_1(1-\tilde g) \big]
\right\}^2.
\label{eq:tVKNBGC}
\end{eqnarray}

\subsection{The covariance of the joint degree distribution of end-nodes of bredges}

The covariance of the joint degree distribution of
end-nodes of random bredges
is given by
\begin{equation}
\Gamma({\rm B}) = 
\widehat{\mathbb{E}}[K K'|{\rm B}]
-  \widetilde{\mathbb{E}}[K|{\rm B}] \
\widetilde{\mathbb{E}}[K'|{\rm B}]
\label{eq:GKK'Bd}
\end{equation}

\noindent
where
$\widehat{\mathbb{E}}[K K'|{\rm B}]$
is the mixed second moment of the joint degree distribution
$\widehat P(k,k'| {\rm B})$
and the mean degree
$\widetilde{\mathbb{E}}[K|{\rm B}]$
of the marginal degree distribution 
is given by Eq. (\ref{eq:tEKB}).
Evaluating the right hand side of Eq. (\ref{eq:GKK'Bd}),
we obtain

\begin{equation}
\Gamma({\rm B}) = - \frac{1}{(1+\tilde g)^2}
\left[ \frac{\langle K^2\rangle}{\langle K\rangle} - 1 - G'_1(1-\tilde g) \right]^2.
\label{eq:GKK'B}
\end{equation}

\noindent
As expected, below the percolation transition, where
$\tilde g=0$, the correlation coefficient is zero. 
In the dilute network regime of $0 < \tilde g \ll 1$, just above the percolation transition, 
\begin{equation}
\Gamma({\rm B}) \simeq -  
\left(  \frac{\langle K^3\rangle}{\langle K\rangle}  - 3 \frac{\langle K^2\rangle}{\langle K\rangle} + 2 \right)^2
\tilde g^2 + \mathcal{O}(\tilde g)^3.
\label{eq:GKK'Bgll1}
\end{equation}

\noindent
In the opposite limit of $\tilde g \rightarrow 1^{-}$ 
the covariance
$\Gamma({\rm B})$ 
converges towards an asymptotic value that depends on the degree distribution. 
It is given by
\begin{equation}
\Gamma({\rm B}) \rightarrow - \frac{1}{4}
\left[ \frac{\langle K^2\rangle}{\langle K\rangle} - 1 -  \frac{ 2 P(2) }{\langle K \rangle}  \right]^2.
\label{eq:GKK'Bgeq1}
\end{equation}

\noindent
The Pearson correlation coefficient for pairs of end-nodes of bredges  
in configuration model networks is given by
\begin{equation}
R({\rm B}) = \frac{\Gamma({\rm B})}{\widetilde{\mathbb{V}}[K|{\rm B}]},
\label{eq:RB}
\end{equation}

\noindent
where $\widetilde{\mathbb{V}}[K|{\rm B}]$
is given by Eq. (\ref{eq:tVKB}).

The covariance of the joint degree distribution of end-nodes of
bredges that reside on the giant component is given by
\begin{equation}
\Gamma({\rm B},{\rm GC}) =
\widehat{\mathbb{E}}[K K'|{\rm B},{\rm GC}]
-  \widetilde{\mathbb{E}}[K|{\rm B},{\rm GC}]  \
\widetilde{\mathbb{E}}[K'|{\rm B},{\rm GC}],
\label{eq:GKK'BGCd}
\end{equation}

\noindent
where $\widehat{\mathbb{E}}[K K'|{\rm B},{\rm GC}]$
is the mixed second moment of $\widehat P(k,k'|{\rm B},{\rm GC})$.
Evaluating the right hand side of Eq. (\ref{eq:GKK'BGCd}),
we obtain
\begin{equation}
\Gamma({\rm B},{\rm GC}) = 
- \frac{1}{4 \tilde g^2}
\left[ \frac{ \langle K^2 \rangle }{ \langle K \rangle } - 1 - G'_1(1-\tilde g) \right]^2.
\label{eq:GKK'BGC}
\end{equation}

\noindent
In the dilute network regime of $0 < \tilde g \ll 1$, just above the percolation transition,
the covariance is given by
\begin{equation}
\Gamma({\rm B},{\rm GC}) \simeq 
- \frac{1}{4}
\left(  \frac{\langle K^3\rangle}{\langle K\rangle}  - 3 \frac{\langle K^2\rangle}{\langle K\rangle} + 2 \right)^2
+ \mathcal{O} \left( \tilde g \right).
\label{eq:GKK'BGCgll1}
\end{equation}

\noindent
In the opposite limit of $\tilde g \rightarrow 1^{-}$ 
the covariance
$\Gamma({\rm B},{\rm GC})$ 
converges towards an asymptotic value that depends on the degree distribution. 
It is given by
\begin{equation}
\Gamma({\rm B},{\rm GC}) \rightarrow - \frac{1}{4}
\left[ \frac{\langle K^2\rangle}{\langle K\rangle} - 1 -  \frac{ 2 P(2) }{\langle K \rangle} \right]^2,
\label{eq:GKK'BGCgeq1}
\end{equation}

\noindent
which is identical to $\Gamma({\rm B})$ in that limit.
The Pearson correlation coefficient for pairs of end-nodes of bredges  
that reside on the giant component is given by
\begin{equation}
R({\rm B},{\rm GC}) = \frac{\Gamma({\rm B},{\rm GC})}{\widetilde{\mathbb{V}}[K|{\rm B},{\rm GC}]},
\label{eq:RBGC}
\end{equation}

\noindent
where $\widetilde{\mathbb{V}}[K|{\rm B},{\rm GC}]$
is given by Eq. (\ref{eq:tVKBGC}).
The end-nodes of non-bredge edges that reside on the giant component 
are actually independent,
as expressed by Eq. (\ref{eq:hPkk'NBGC2}), and in particular they exhibit no degree-degree correlations. 
Therefore, 
$R({\rm NB},{\rm GC}) = 0$.

\end{widetext}

\section{Applications to specific network models}

Here we apply the approach presented above 
to several examples of configuration model networks
with given degree distribution.
More specifically, we consider the cases of
the Poisson degree distribution (ER networks),
the exponential degree distribution 
and the power-law degree distribution (scale-free networks).

\subsection{Erd{\H o}s-R\'enyi networks}

The ER network is
a random network in which
each pair of nodes 
is connected
with probability
$p$
\cite{Erdos1959,Erdos1960,Erdos1961}.
The mean degree of an ER network is
$c=(N-1)p$,
where $N$ is the network size,
and the degree distribution is a
Poisson distribution of the form
\cite{Bollobas2001}
\begin{equation}
P(k) = \frac{e^{-c} c^k}{k!}.
\label{eq:poisson}
\end{equation}

\noindent
Since it exhibits no correlations, the ER network is a configuration model
network with a Poisson degree distribution.
Moreover, it is a maximum entropy network under the condition
that the mean degree 
$\langle K \rangle = c$  
is constrained.
Asymptotic ER
networks exhibit a percolation transition at $c=1$, such that for
$c<1$
the network consists only of finite components,
which exhibit tree topologies.
For  
$c>1$
a giant component emerges, coexisting with the finite components. 
At a higher value of the connectivity, namely at 
$c = \ln N$, 
there is a second transition, above which
the giant component encompasses 
the entire network. 

ER networks exhibit a special property, 
resulting from the Poisson degree distribution
[Eq. (\ref{eq:poisson})], 
which satisfies
$\widetilde P(k) = P(k-1)$,
where ${\widetilde P}(k)$ is given by Eq. (\ref{eq:tPk}).
This implies that for the Poisson distribution 
the two generating functions 
are identical, namely $G_1(x)=G_0(x)$.
Using Eqs. (\ref{eq:tg}) and (\ref{eq:g}) we obtain that for ER networks
${\tilde g} = g$. 
Carrying out the summations in Eqs. (\ref{eq:G0}) and (\ref{eq:G1})
with $P(k)$ given by Eq. (\ref{eq:poisson}), 
one obtains
$G_0(x)=G_1(x)=e^{-(1-x)c}$.
Inserting this result in Eq. (\ref{eq:g}),
it is found that $g$ satisfies the equation
$1-g =  e^{-gc}$
\cite{Bollobas2001}.
Solving for the probability $g$ as a function of the mean degree, $c$, 
one obtains
\begin{equation}
g = \tilde g = 1 + \frac{W(-c e^{-c})}{c},
\label{eq:g(c)}
\end{equation}

\noindent
where $W(x)$ is the Lambert $W$ function
\cite{Olver2010}.


\begin{figure}[b!]
\includegraphics[width=7.5cm]{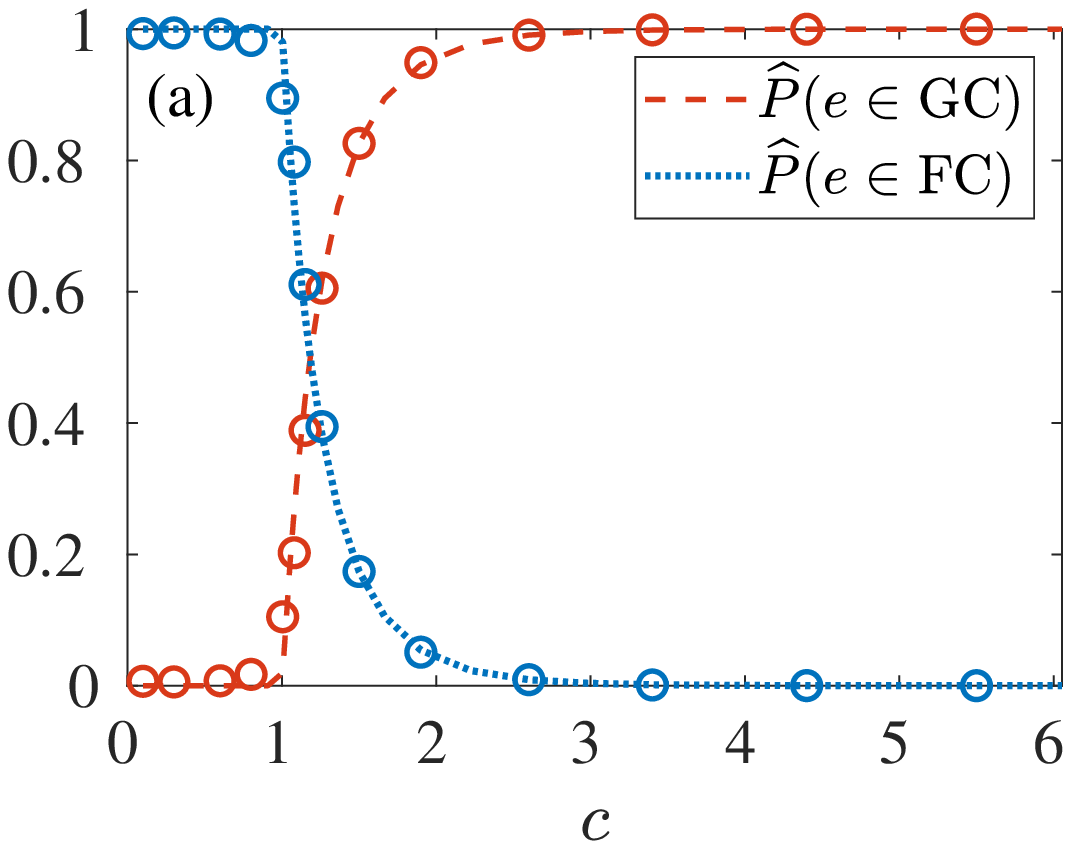} 
\\
\includegraphics[width=7.5cm]{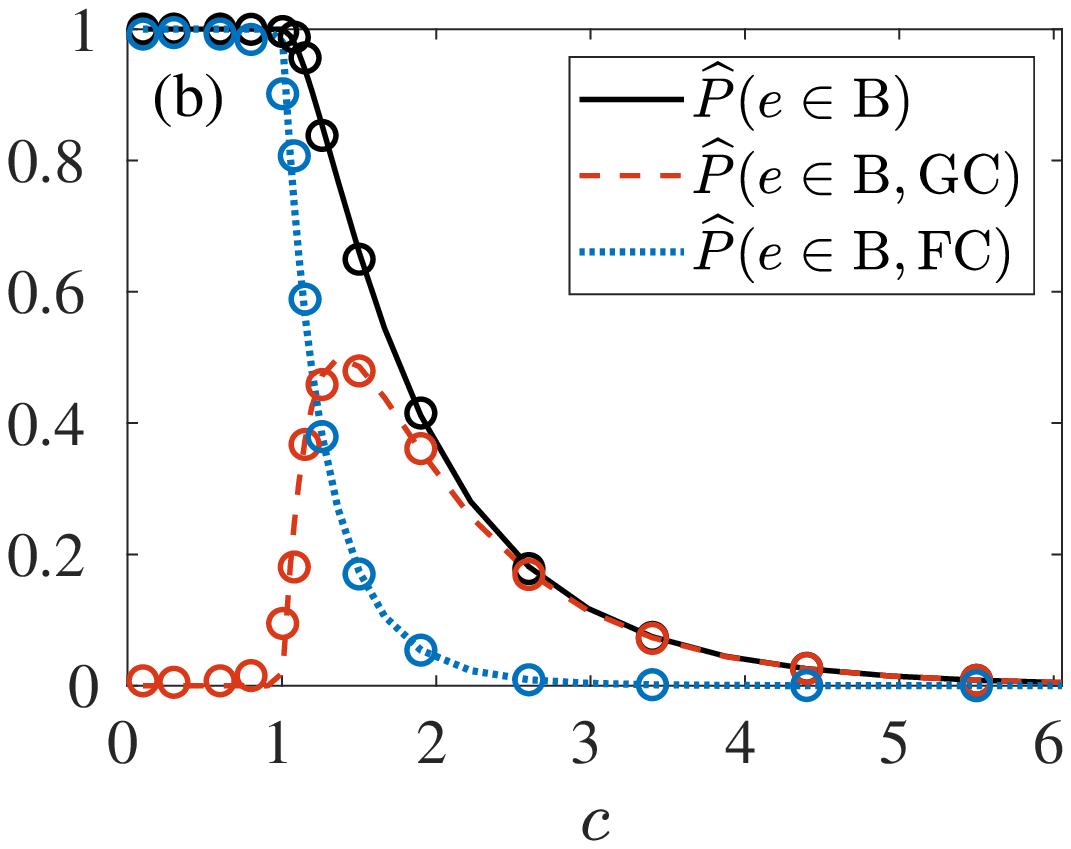} 
\caption{
(Color online)
(a) The probability $\widehat P(e \in {\rm GC})$ (dashed line),
that a randomly sampled edge in an ER network resides on the giant component,
as a function of the mean degree $c=\langle K \rangle$,
obtained from Eq. (\ref{eq:hPeGC});
The complementary probability $\widehat P(e \in {\rm FC})$ (dotted line) 
that a randomly sampled edge resides on
one of the finite components 
is also shown.
(b) The probability
$\widehat P(e \in {\rm B})$ (solid line)
that a randomly sampled edge in an ER network is a bredge,
as a function of the mean degree $c$, 
obtained from Eq. (\ref{eq:hPeB});
The probability 
$\widehat P(e \in {\rm B})$ 
is equal to the sum of two components:
the probability
$\widehat P(e \in {\rm B} ,   {\rm GC})$
(dashed line)
that a randomly sampled edge 
is a bredge that resides on the giant component,
and the probability
$\widehat P(e \in {\rm B} ,  {\rm FC})$
(dotted line)
that a randomly sampled edge 
is a bredge that resides on one of the finite
components.
The analytical results
are in excellent agreement with 
the results of
computer simulations
(circles),
performed for an ensemble of ER networks
of $N=10^4$ nodes.
}
\label{fig:5}
\end{figure}

In Fig. 5(a) we present analytical results for the
probability $\widehat P(e \in {\rm GC})$ (dashed line),
that a randomly sampled edge in an ER network resides on the giant component,
as a function of the mean degree $c$.
These results are obtained 
by inserting $\tilde g$ from Eq. (\ref{eq:g(c)})
into Eq. (\ref{eq:hPeGC}).
We also present 
the complementary probability 
$\widehat P(e \in {\rm FC})$ (dotted line) 
that a randomly sampled edge resides on
one of the finite components.
In Fig. 5(b) we present analytical reults for the
probability
$\widehat P(e \in {\rm B})$ (solid line)
that a randomly sampled edge in an ER network is a bredge
as a function of the mean degree $c$. 
The probability
$\widehat P(e \in {\rm B})$
can be expressed as a sum of two components:
the probability
$\widehat P(e \in {\rm B} , {\rm GC}) = \widehat P(e \in {\rm B}|{\rm GC}) \widehat P(e \in {\rm GC})$ 
(dashed line)
that a randomly sampled edge 
is a bredge that resides on the giant component,
and the probability
$\widehat P(e \in {\rm B} , {\rm FC})  = \widehat P(e \in {\rm FC})$ 
(dotted line)
that a randomly sampled edge is a bredge that resides on one of the finite
components.
The analytical results
are in excellent agreement with 
the results of
computer simulations
(circles),
performed for an ensemble of ER networks
of $N=10^4$ nodes.


\begin{figure}
\includegraphics[width=7.5cm]{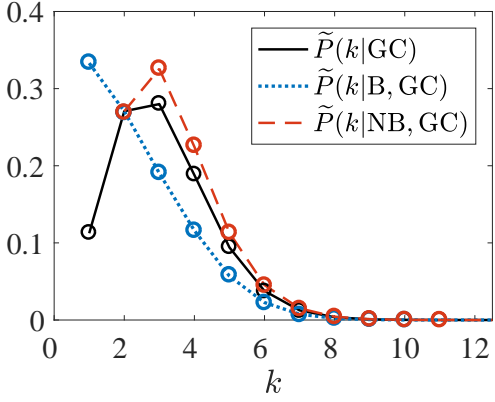}
\caption{
(Color online)
Analytical results for the marginal degree distribution 
$\widetilde P(k|{\rm GC})$ (solid line)
of end-nodes of randomly sampled edges,
the marginal degree distribution
$\widetilde P(k|{\rm B},{\rm GC})$ (dotted line)
of end-nodes of randomly sampled bredges 
and the marginal degree distribution
$\widetilde P(k|{\rm NB},{\rm GC})$ (dashed line)
of randomly sampled non-bredge edges 
on the giant component
of an ER network with mean degree
$c=2$.
The analytical results are in excellent agreement with the 
results obtained from computer simulations 
(circles).
}
\label{fig:6}
\end{figure}

In Fig. 6 we present analytical results for the marginal degree distribution
$\widetilde P(k|{\rm GC})$
of end-nodes of edges on the giant component of an ER network (solid line),
obtained by inserting $\tilde g$ from Eq. (\ref{eq:g(c)}) 
into Eq. (\ref{eq:tPkGC}).
We also present the marginal degree distribution
$\widetilde P(k|{\rm B},{\rm GC})$
of end-nodes of bredges on the giant component (dotted line),
obtained by inserting $\tilde g$ from Eq. (\ref{eq:g(c)}) into Eq. (\ref{eq:tPkBGC})
and for the marginal degree distribution
$\widetilde P(k|{\rm NB},{\rm GC})$ of
non-bredge edges that reside on the giant component (dashed line),
obtained by inserting $\tilde g$ from Eq. (\ref{eq:g(c)}) into Eq. (\ref{eq:tPkNBGC}).
The analytical results are in excellent agreement with the corresponding 
results obtained from computer simulations (circles).
It is found that the marginal degree distribution of the end-nodes of bredges
decreases monotonically as a function of $k$, while the marginal degree distribution
of the non-bredge edges exhibits a peak. 
Overall, the degrees of end-nodes of non-bredge edges tend to be higher
than the degrees of end-nodes of bredges.


\begin{figure}
\includegraphics[width=7.5cm]{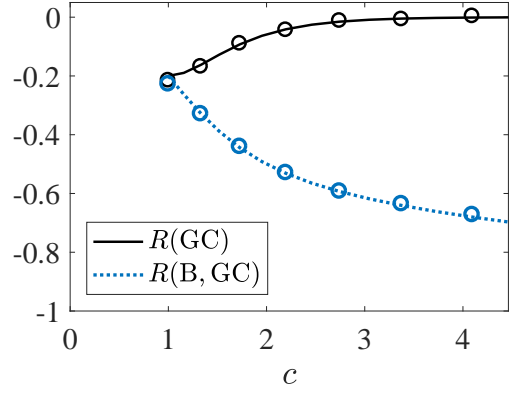}
\caption{
(Color online)
Analytical results for the correlation coefficient 
$R({\rm GC})$ (solid line)
between the degrees
$k$ and $k'$ of end-nodes of randomly sampled edges
and the correlation coefficient
$R({\rm B},{\rm GC})$ (dotted line)
between the end-nodes
of randomly sampled bredges 
that reside on the giant component of an ER network,
as a function of the mean degree $c$.
The analytical results are in excellent agreement with the results obtained from
computer simulations (circles).
The correlations, which are concentrated on the bredges, are negative
and become stronger as $c$ is increased.
Since the fraction of bredges on the giant component
is a decreasing function of $c$,
the correlation coefficient over all the edges on the giant component
decreases (in absolute value) as $c$ is increased.
}
\label{fig:7}
\end{figure}

In Fig. 7 we present 
analytical results for the correlation coefficient
$R({\rm GC})$
between the degrees of pairs of end-nodes of edges on the giant component
of an ER network as a function of the mean degree $c$ (solid line),
obtained by inserting $\tilde g$ from Eq. (\ref{eq:g(c)}) into Eq. (\ref{eq:RGC}).
We also present
the correlation coefficient 
$R({\rm B},{\rm GC})$
between the degrees of end-nodes of bredges that reside 
on the giant component of an ER network (dotted line),
obtained by inserting $\tilde g$ from Eq. (\ref{eq:g(c)}) into Eq. (\ref{eq:RBGC}).
The analytical results are in excellent agreement with the results obtained from
computer simulations (circles).

\subsection{Configuration model networks with exponential degree distributions}

Consider a configuration model network with an exponential degree
distribution of the form $P(k) \sim e^{- \alpha k}$,
where $k_{\rm min} \le k \le k_{\rm max}$.
In case that $k_{\rm min} \ge 2$ one can show that
$g = \tilde g =1$ and there are no bredges.
Here we consider the case of
$k_{\rm min}=1$ 
and 
$k_{\rm max} = \infty$.
In this case it is convenient to parametrize the degree distribution 
using the mean degree $c$
in the form
\begin{equation}
P(k) = \frac{1}{c-1} \left( \frac{c-1}{c} \right)^k.
\label{eq:exp}
\end{equation}

\noindent
In order to find the properties of bredges in such networks,
we first calculate the parameters $\tilde g$ and $g$. 
Inserting the exponential degree distribution of 
Eq. (\ref{eq:exp}) 
into the generating function
$G_1(x)$, given by Eq. (\ref{eq:G1}),
we obtain
$G_1(x) =  [ c - (c-1) x ]^{-2}$.
Inserting the above expression of $G_1(x)$ 
into 
Eq. (\ref{eq:tg}) and
solving for $\tilde g$, 
we find that
for $c > 3/2$
there is a non-trivial solution of the form
\begin{equation}
\tilde g = \frac{1}{2} \left[ \frac{c-3}{c-1} +  \sqrt{ \frac{c+3}{c-1} } \right].
\label{eq:tgexp}
\end{equation}

\noindent
Inserting the exponential degree distribution of Eq. (\ref{eq:exp}) into
Eq. (\ref{eq:G0}), we obtain
$G_0(x) =  {x}/{ [c - (c-1)x] }$.
Inserting $\tilde g$ 
from Eq. (\ref{eq:tgexp})
and the above expression of
$G_0(1-\tilde g)$ into
Eq. (\ref{eq:g}),
we find that for 
$c > 3/2$
\begin{equation}
g = 
\frac{ c }{ 2 (c-1) }
\left[ 3 - \sqrt{ \frac{c+3}{c-1} } \right].
\label{eq:gexp}
\end{equation}

\noindent
Thus, it is found that the configuration model network with an exponential
degree distribution exhibits a percolation transition at $c_0 = 3/2$.


\begin{figure}
\includegraphics[width=7.5cm]{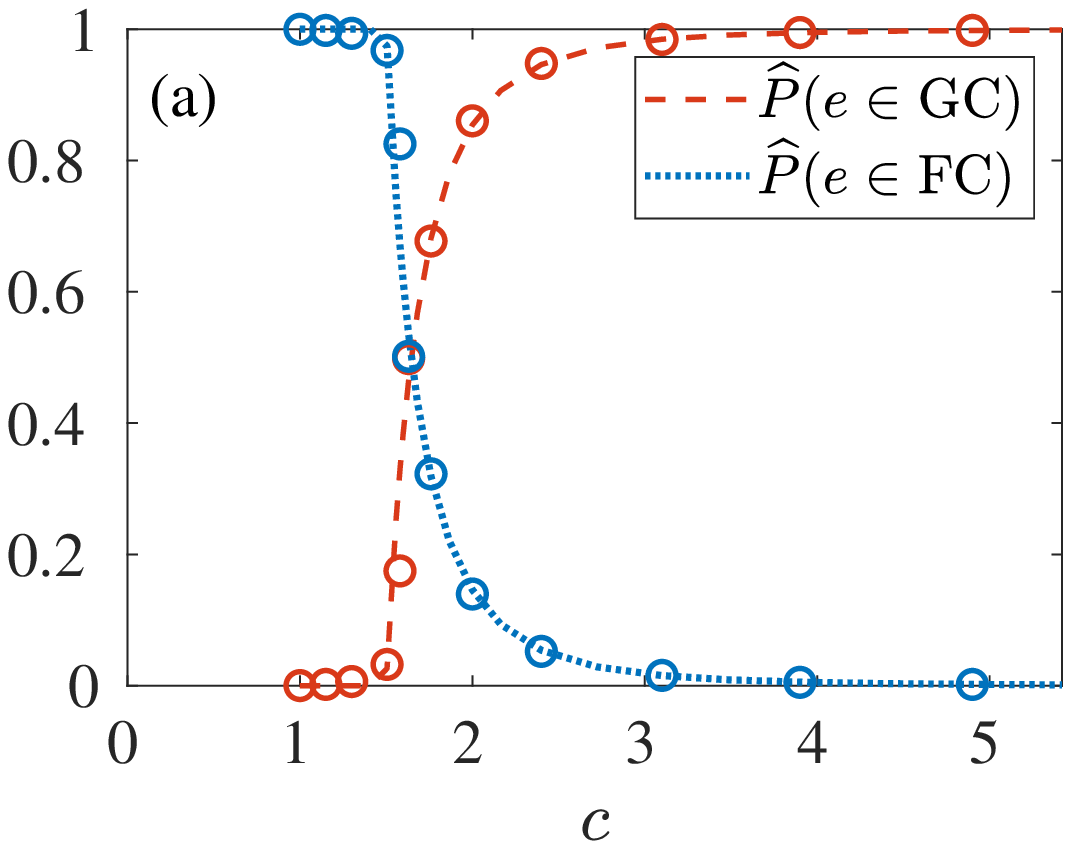} 
\\
\includegraphics[width=7.5cm]{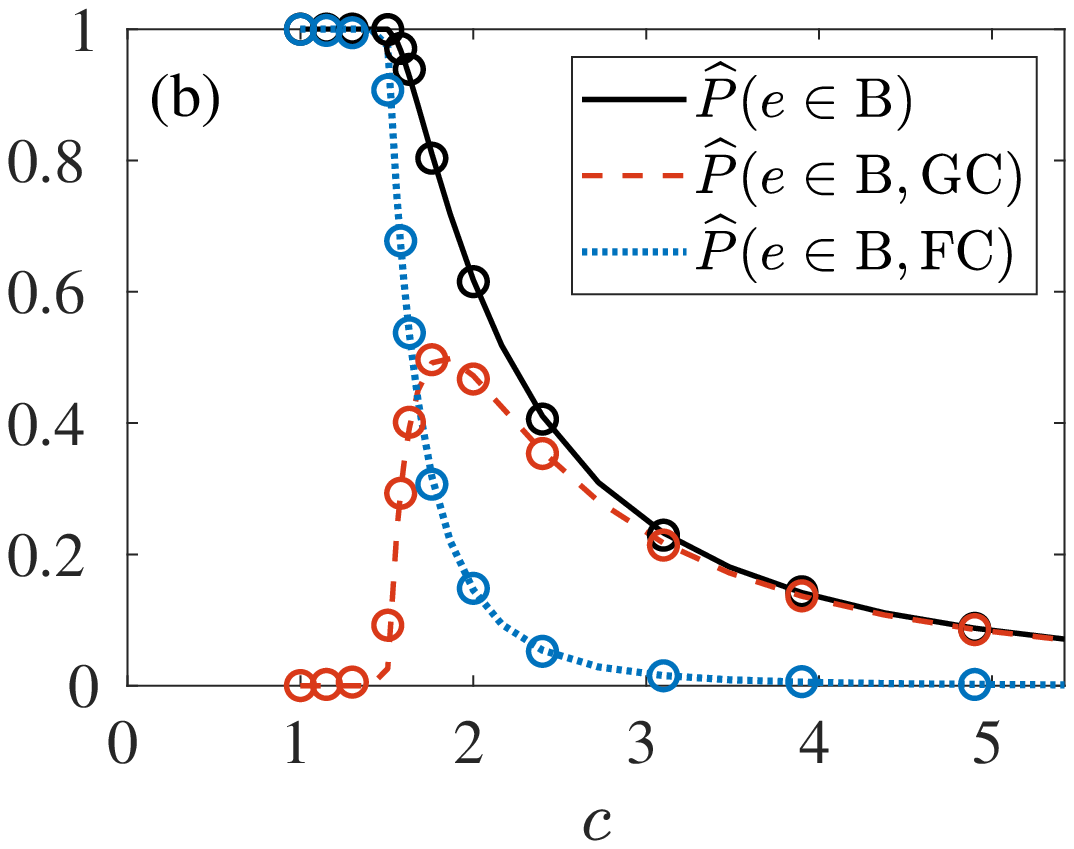} 
\caption{
(Color online)
(a) Analytical results for the probability 
$\widehat P(e \in {\rm GC})$ (dashed line)
that a randomly sampled edge 
in a configuration model network with an exponential degree distribution 
resides on the giant component,
as a function of the mean degree $c$;
The complementary probability 
$P(e \in {\rm FC})$ (dotted line)
that a randomly sampled edge resides on
one of the finite tree components  
is also shown.
(b) Analytical results for the probability
$\widehat P(e \in {\rm B})$ (solid line)
that a randomly sampled edge 
is a bredge,
as a function of the mean degree $c$,
obtained from Eq. (\ref{eq:hPeB});
The probability
$P(e \in {\rm B})$
is equal to the sum of two components:
the probability
$\widehat P(e \in {\rm B} , {\rm GC})$
(dashed line),
that a randomly sampled edge 
is a bredge that resides on the giant component
and the probability
$\widehat P(e \in {\rm B} ,   {\rm FC})$
that a randomly sampled edge is a bredge that resides on one of the finite
components.
The analytical results
are in excellent agreement with 
the results of
computer simulations
(circles),
performed for an ensemble of configuration model networks
of $N=10^4$ nodes.
}
\label{fig:8}
\end{figure}

In Fig. 8(a) we present the 
probability $\widehat P(e \in {\rm GC})$ (dashed line),
that a random edge in a configuration model network with an exponential degree distribution
resides on the giant component,
obtained from Eq. (\ref{eq:hPeGC}),
and the probability $\widehat P(e \in {\rm FC})$ (dotted line) that a random edge resides on
one of the finite components,
as a function of the mean degree $c$.
In Fig. 8(b) we present the
probability
$\widehat P(e \in {\rm B})$ 
that a random edge in a configuration model network with an exponential degree distribution is a bredge
(solid line),
as a function of $c$, 
obtained from Eq. (\ref{eq:hPeB}).
We also present
the probability
$\widehat P(e \in {\rm B} ,  {\rm GC})$ (dashed line)
that a randomly selected edge in the network
is a bredge that resides in the giant component
and the probability
$\widehat P(e \in {\rm B} ,   {\rm FC})$ (dotted line)
that a randomly selected edge in the network is a bredge that resides in one of the finite
components.
The analytical results
are found to be in excellent agreement with 
the results of
computer simulations
(circles),
performed for an ensemble of configuration model networks
of $N=10^4$ nodes.


\begin{figure}[h!]
\includegraphics[width=7.5cm]{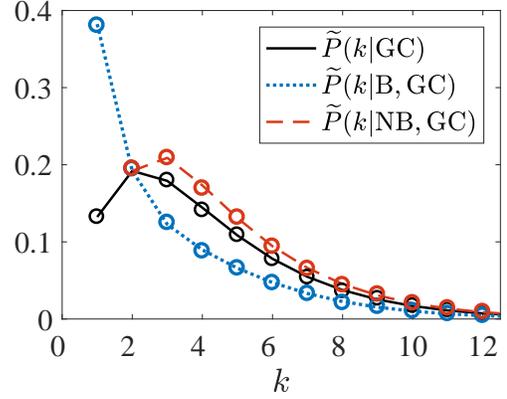}
\caption{
(Color online)
Analytical results for the marginal degree distribution 
$\widetilde P(k|{\rm GC})$ (solid line)
of end-nodes of randomly sampled edges,
the marginal degree distribution
$\widetilde P(k|{\rm B},{\rm GC})$ (dotted line)
of end-nodes of randomly sampled bredges, 
and the marginal degree distribution
$\widetilde P(k|{\rm NB},{\rm GC})$ (dashed line)
of randomly sampled non-bredge edges,
on the giant component
of a configuration model network 
with an exponential degree distribution
and mean degree $c=2.5$.
The analytical results are in excellent agreement with the 
results obtained from computer simulations 
(circles).
}
\label{fig:9}
\end{figure}


\begin{figure}[h!]
\includegraphics[width=7.5cm]{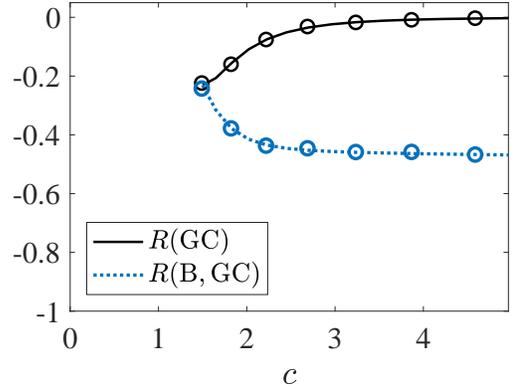}
\caption{
(Color online)
Analytical results for the correlation coefficient 
$R({\rm GC})$ (solid line) 
between the degrees
$k$ and $k'$ of end-nodes of edges and
the correlation coefficient
$R({\rm B},{\rm GC})$ (dotted line)
between the end-nodes
of bredges that reside on
the giant component of a configuration model network
with an exponential degree distribution,
as a function of the mean degree $c$.
The analytical results are in excellent agreement with the results obtained from
computer simulations (circles).
}
\label{fig:10}
\end{figure}

In Fig. 9 we present
analytical results for the marginal degree distribution 
$\widetilde P(k|{\rm GC})$
of end-nodes of randomly selected edges
(solid line),
the marginal degree distribution
$\widetilde P(k|{\rm B},{\rm GC})$
of end-nodes of bredges 
(dotted line)
and the marginal degree distribution
$\widetilde P(k|{\rm NB},{\rm GC})$
of non-bredge edges 
(dashed line) 
on the giant component
of a configuration model network
with an exponential degree distribution.
The analytical results are in excellent agreement with the 
results obtained from computer simulations (circles).

In Fig. 10 we present
analytical results for the correlation coefficients 
$R({\rm GC})$ 
and 
$R({\rm B},{\rm GC})$
between the degrees
$k$ and $k'$ of the end-nodes of edges
that reside on the giant component (solid line)
and bredges that reside on the giant component
(dotted line), respectively,
as a function of the mean degree $c$
in configuration model networks with exponential degree distributions.
The analytical results are in excellent agreement with the results obtained from
computer simulations (circles).


\begin{figure}[h!]
\includegraphics[width=7cm]{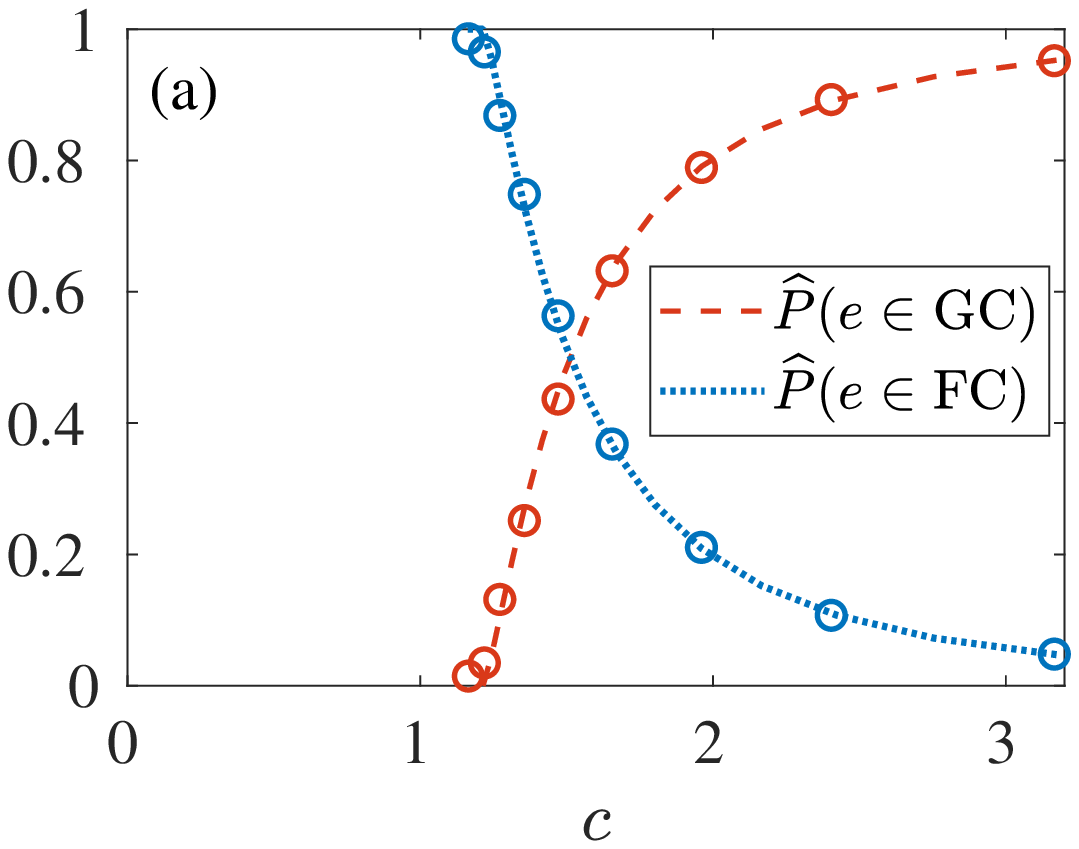} 
\\
\includegraphics[width=7cm]{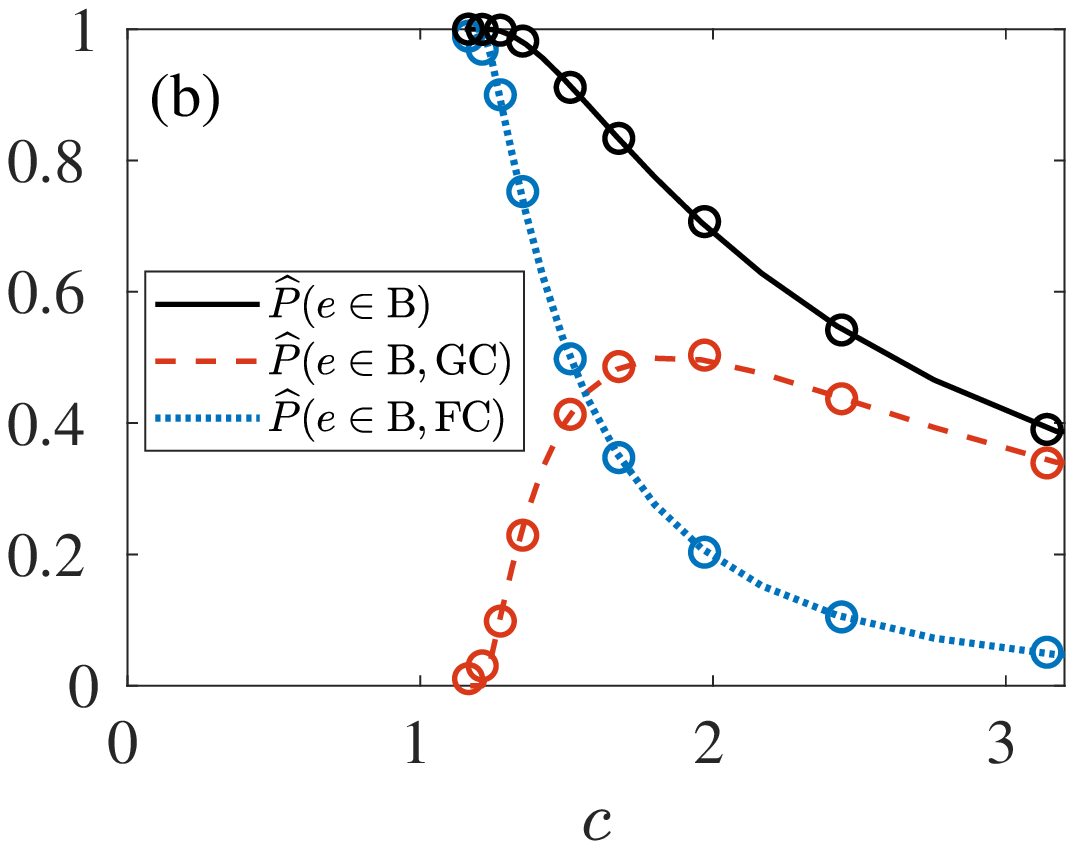} 
\caption{
(Color online)
(a) Analytical results for the probability 
$\widehat P(e \in {\rm GC})$  (dashed line)
that a randomly sampled edge 
in a configuration model network with a power-law degree distribution
resides on the giant component,
as a function of the mean degree $c$.
The complementary probability $\widehat P(e \in {\rm FC})$ (dotted line)
that a randomly sampled edge resides on
one of the finite tree components is also shown;
(b) Analytical results for the probability
$\widehat P(e \in {\rm B})$ (solid line)
that a random edge 
is a bredge,
as a function of the mean degree $c$.
The probability 
$\widehat P(e \in {\rm B})$
is equal to the sum of two components: 
the probability
$\widehat P(e \in {\rm B} , {\rm GC})$ (dashed line)
that a randomly sampled edge 
is a bredge that resides on the giant component
and the probability
$\widehat P(e \in {\rm B} , {\rm FC})$ (dotted line)
that a randomly sampled edge is a bredge that resides on one of the finite
components.
The analytical results
are in excellent agreement with 
the results of
computer simulations
(circles),
performed for networks
of $N=10^4$ nodes.
}
\label{fig:11}
\end{figure}

\subsection{Configuration model networks with power-law degree distributions}

Consider a configuration model network with a power-law degree distribution
of the form 
$P(k) \sim k^{-\gamma}$,
where 
$k_{\rm min} \le k \le k_{\rm max}$.
For $\gamma \le 2$ the mean degree diverges in the limit of 
$k_{\rm max} \rightarrow \infty$.
For $2 < \gamma \le 3$ the mean degree is bounded while the second moment 
diverges.
For $\gamma > 3$ both moments are bounded.
Here we focus on the case of $\gamma > 2$, in which the
mean degree, 
$\langle K \rangle$, 
is bounded even for 
$k_{\rm max} \rightarrow \infty$.
We choose $k_{\rm min}=1$, for which there is a coexistence
phase of the giant component and the finite tree components
and $k_{\rm max}=100$.
The normalized degree distribution is given by
\begin{equation}
P(k) = A(\gamma,k_{\rm max}) \  k^{-\gamma},  
\label{eq:PLnorm}
\end{equation}

\noindent
where the normalization factor is
$A(\gamma,k_{\rm max}) = [ \zeta(\gamma) - \zeta(\gamma,k_{\rm max}+1) ]^{-1}$,
the function
$\zeta(\gamma,k)$ is the Hurwitz zeta function
and
$\zeta(\gamma) = \zeta(\gamma,1)$ is the Riemann zeta function 
\cite{Olver2010}.
The mean degree is given by
$\langle K \rangle = A(\gamma,k_{\rm max})/A(\gamma-1,k_{\rm max})$
and the second moment of the degree distribution is given by
$\langle K^2 \rangle = A(\gamma,k_{\rm max})/A(\gamma-2,k_{\rm max})$.
Inserting the degree distribution of Eq. (\ref{eq:PLnorm})
into Eqs. 
(\ref{eq:G0})
and
(\ref{eq:G1})
we obtain
\begin{equation}
G_0(x) = 
A(\gamma,k_{\rm max})
\left[ {\rm Li}_{\gamma}(x) - x^{k_{\rm max}+1}  \Phi(x,\gamma,k_{\rm max}+1)  \right],
\end{equation}

\noindent
and
\begin{eqnarray}
x G_1(x) & = &  
  A(\gamma-1,k_{\rm max})  
\Big[ {\rm Li}_{\gamma-1}(x)  \nonumber\\
& &- x^{k_{\rm max}+1} \Phi(x,\gamma-1,k_{\rm max}+1)  \Big],
\end{eqnarray}

\noindent
where 
$\Phi(x,\gamma,k)$ is the Lerch transcendent and
${\rm Li}_{\gamma}(x)$ is the polylogarithm function
\cite{Gradshteyn2000}.
The values of the parameters $\tilde g$ and $g$ are determined by
Eqs. (\ref{eq:tg}) and (\ref{eq:g}).
Unlike the ER network and the configuration model network
with an exponential degree distribution, here we do not have
closed form analytical expressions for $g$ and $\tilde g$.
However, using the expressions above
for $G_0(x)$ and $G_1(x)$, 
the values of $g$ and $\tilde g$ can be easily obtained
from a numerical solution of 
Eqs. (\ref{eq:tg}) and (\ref{eq:g}).
Using the 
Molloy-Reed criterion
\cite{Molloy1995,Molloy1998},
we find that
for $k_{\rm max}=100$
the percolation threshold is
$c_0 \simeq 1.219$,
where
$\gamma_0 \simeq 3.378$.

In Fig. 11(a) we present the
probability $\widehat P(e \in {\rm GC})$  
that a random edge in a configuration model network with a power-law degree distribution
resides on the giant component (dashed line),
obtained from Eq. (\ref{eq:hPeGC}), as a function of $c$.
We also present the complementary
probability $\widehat P(e \in {\rm FC})$ that a random edge resides on
one of the finite components (dotted line).
In Fig. 11(b) we present the
probability
$\widehat P(e \in {\rm B})$ 
that a random edge in a configuration model network with a power-law degree distribution is a bredge
(solid line),
as a function of $c$, 
obtained from Eq. (\ref{eq:hPeB}).
We also present 
the probability
$\widehat P(e \in {\rm B} , {\rm GC})$ (dashed line)
that a randomly selected edge 
is a bredge that resides on the giant component
and the probability
$\widehat P(e \in {\rm B} , {\rm FC})$ (dotted line)
that a randomly selected edge is a bredge that resides on one of the finite tree
components.
The analytical results
are in excellent agreement with 
the results of
computer simulations
(circles),
performed for an ensemble of configuration model networks
of $N=10^4$ nodes.


\begin{figure}
\includegraphics[width=7.5cm]{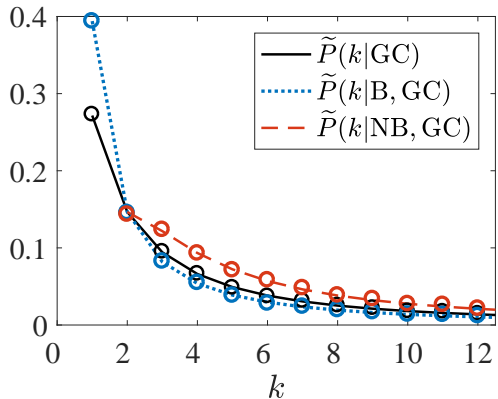}
\caption{
(Color online)
Analytical results for the marginal degree distribution 
$\widetilde P(k|{\rm GC})$
of end-nodes of randomly selected edges
(solid line),
the marginal degree distribution
$\widetilde P(k|{\rm B},{\rm GC})$
of end-nodes of bredges 
(dotted line)
and the marginal degree distribution
$\widetilde P(k|{\rm NB},{\rm GC})$
of end-nodes of non-bredge edges 
(dashed line) 
on the giant component
of a configuration model network
with a power-law degree distribution
with an exponent $\gamma=2.5$ and mean degree $c=1.54$.
The analytical results are in excellent agreement with the 
results obtained from computer simulations
(circles).
}
\label{fig:12}
\end{figure}

\begin{figure}
\includegraphics[width=7.5cm]{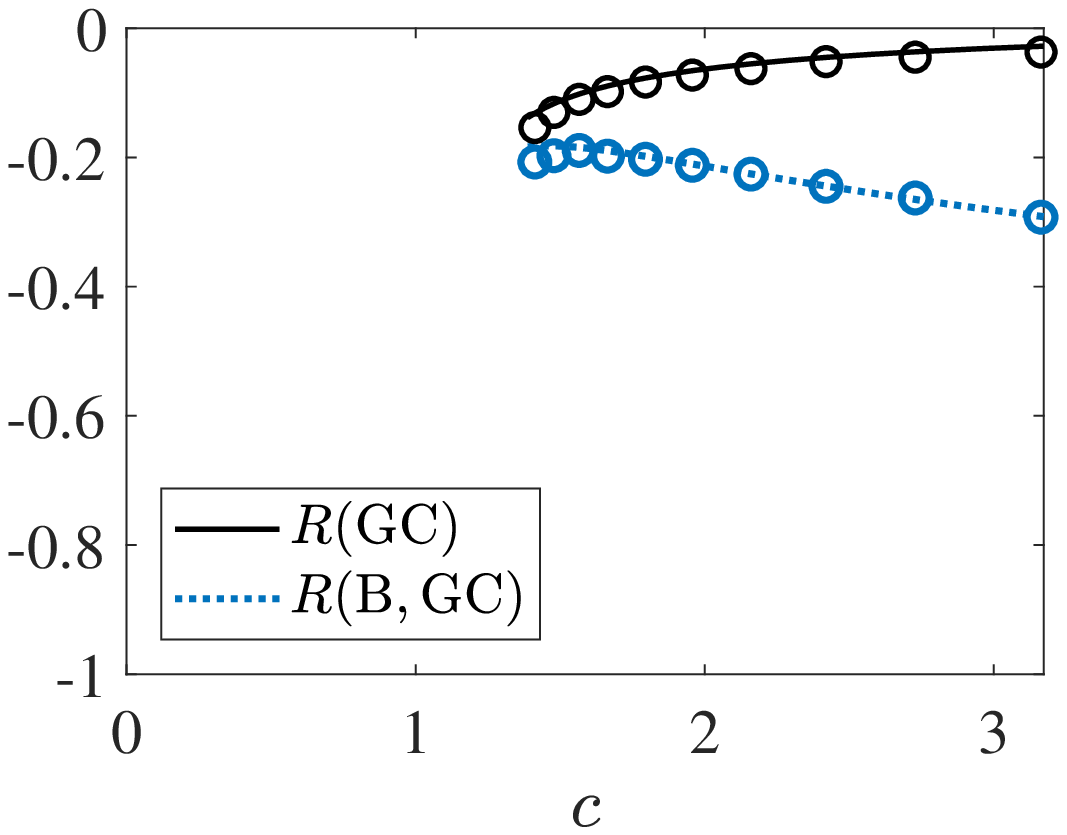}
\caption{
(Color online)
Analytical results for the correlation coefficient  
$R({\rm GC})$, 
between the degrees
$k$ and $k'$ of end-nodes of edges
(solid line) and
the correlation coefficient
$R({\rm B},{\rm GC})$ 
between the end-nodes
of bredges 
(dotted line) 
that reside on
the giant component of a configuration model network
with a power-law degree distribution,
as a function of the mean degree $c$.
The analytical results are in excellent agreement with the results obtained from
computer simulations (circles), except for the dilute network regime just above the
percolation transition.
In this regime the giant component is small and its size fluctuates between 
different network instances. 
The data points in this regime were averaged over 100 network instances,
while all the other data points were averaged over 20 network instances.
}
\label{fig:13}
\end{figure}

In Fig. 12 we present analytical results for the marginal degree distribution 
$\widetilde P(k|{\rm GC})$
of end-nodes of randomly selected edges
(solid line),
the marginal degree distribution
$\widetilde P(k|{\rm B},{\rm GC})$
of end-nodes of bredges 
(dotted line)
and the marginal degree distribution
$\widetilde P(k|{\rm NB},{\rm GC})$
of non-bredge edges 
(dashed line) 
on the giant component
of a configuration model network
with a power-law degree distribution.
The analytical results are in excellent agreement with the 
results obtained from computer simulations (circles).

In Fig. 13 we present analytical
results for the correlation coefficient
$R({\rm GC})$, 
between the degrees
$k$ and $k'$ of end-nodes of edges
(solid line) and
the correlation coefficient
$R({\rm B},{\rm GC})$ 
between the end-nodes
of bredges 
(dotted line) 
that reside on
the giant component of a configuration model network
with a power-law degree distribution,
as a function of the mean degree $c$.
The analytical results are in excellent agreement with the results obtained from
computer simulations (circles) except for the dilute network regime where
there are noticeable deviations due to finite size effects.

In the case of infinite networks, one may consider the
limit of $k_{\rm max} \rightarrow \infty$.
In this limit
the expression for the degree distribution is simplified to
$P(k)=k^{-\gamma}/\zeta(\gamma)$.  
For $\gamma > 2$ the mean degree is given by
$\langle K \rangle = \zeta(\gamma-1)/\zeta(\gamma)$
and for $\gamma > 3$ 
the second moment is given by
$\langle K^2 \rangle = \zeta(\gamma-2)/\zeta(\gamma)$.
The generating functions are simplified to
$G_0(x) = {\rm Li}_{\gamma}(x)/\zeta(\gamma)$
and
$x G_1(x) = {\rm Li}_{\gamma-1}(x)/\zeta(\gamma-1)$.
Using the 
Molloy-Reed criterion
\cite{Molloy1995,Molloy1998},
we find that
for $k_{\rm max} \rightarrow \infty$
the percolation threshold is
$c_0 \simeq 1.196$,
where
$\gamma_0 \simeq 3.478$.

\section{Discussion}

Transportation, communication and many other networks consist of a single
connected component, in which there is at least one path connecting any
pair of nodes. This property is essential for the functionality of these networks.
The failure of a node or an edge
disconnects the paths that go through the failed node/edge.
In case that the failed node is an AP or the failed edge is a bredge,
the disconnected paths have no substitute.
As a result, a whole patch of nodes becomes disconnected
from the rest of the network.
Networks that do not include any APs and bredges are called biconnected networks
\cite{Newman2008,Dufresne2013}.
In such networks, any node $i$ is connected to any other node $j$  
by at least two non-overlapping paths.
While biconnected networks are resilient to the deletion of a single node
or a single edge,
they are still vulnerable to multiple node/edge deletions. 
This is due to the fact that
the deletion of a node/edge may turn other nodes into APs
and other edges into bredges. 
Their subsequent deletion would disconnect other nodes from the rest of the network.
The properties of APs and bredges are utilized in optimized algorithms of network dismantling
\cite{Braunstein2016,Zdeborova2016,Wandelt2018,Ren2019}.
The first stage of these dismantling processes is the decycling stage in
which one node is deleted in each cycle, transforming the network into a tree
network. In tree networks all the nodes of degrees $k \ge 2$ are APs 
and all the edges are bredges. 
Thus the deletion of such nodes/edges efficiently breaks the network into many
small components.

The properties of bredges in a wide range of
real-world empirical networks were recently studied
\cite{Wu2018}.
The fraction of bredges in each network was calculated
using an algorithm based on depth-first search.
An ensemble of configuration model networks, whose degree 
distribution coincides with the degree sequence of the empirical network,
was generated using degree-preserving randomization.
The fraction of bredges in each ensemble was calculated
both numerically and using a generating function formalism.
It was found that the fraction of bredges in the
randomized ensembles is very similar to their fraction in the 
corresponding empirical networks. This indicates that the 
information about the number of bredges is captured in the
degree distribution. Therefore, correlations and other structural
properties that distinguish an empirical network
from the corresponding configuration model network were
found to have little effect on the number of bredges.

The edges in a network can be considered as the building blocks of  
paths connecting pairs of nodes.
Pairs of nodes that reside on the same network component may 
be connected to each other by multiple paths.
Among the paths connecting a pair of nodes $i$ and $j$, the shortest
paths are of particular importance
because they are likely to provide the fastest and
strongest interactions. 
The statistical properties of the shortest paths are captured by
the distribution of shortest path lengths (DSPL).
The DSPL can be used to characterize the large scale structure
of the network, in analogy to the degree destribution which 
is used to characterize the local structure.
Central measures of the DSPL
such as the mean distance 
\cite{Bollobas2001,Chung2003,Fronczak2004,Durrett2007}
and extremal measures such
as the diameter 
\cite{Hartmann2017}
were studied. 
However, apart from a few studies 
\cite{Newman2001,Dorogovtsev2003,Blondel2007,Hofstad2008,Esker2008,Shao2008,Shao2009}
the DSPL
has not attracted nearly as much attention as the degree distribution. 
Recently, an analytical approach was developed for calculating the DSPL in the  
(ER) network 
\cite{Katzav2015}, 
followed by more general formulations
that apply to configuration model networks 
\cite{Nitzan2016,Melnik2016,Katzav2018},
to modular networks 
\cite{Asher2020}
and to networks that form by kinetic growth processes
\cite{Steinbock2017,Steinbock2019a,Steinbock2019b}.

The importance of a given edge $e$ in a network may be quantified by its
betweeness centrality, which is the number 
of pairs of nodes $i$ and $j$, such that
of shortest paths between
them pass through $e$
\cite{Freeman1977,Goh2003}.
In general, the calculation of the betweeness centrality of an edge
cannot be done locally. It involves the calculation of the 
shortest paths between all the pairs of nodes in the network, which requires
access to the structure of the whole network
\cite{Brandes2001}.
However, in case that an edge $e$ is a bredge, one can easily
obtain its betweeness centrality.
Consider a bredge $e$ that resides on the giant component 
whose size is $N_{\rm GC}$.
If the deletion of $e$ detaches a tree component of size $N_{\rm FC}$
from the giant component,
the betweeness centrality of $e$ is given by
$\beta_{e} = N_{\rm FC}(N_{\rm GC} - N_{\rm FC})$.

The damage exerted on a network upon deletion of a bredge
can be evaluated using a centrality measure called bridgeness
\cite{Wu2018}.
The bridgeness of a bredge $e$ 
that resides on the giant component
is defined as the number of nodes disconnected
from the giant component upon deletion of $e$.
The bridgeness of bredges that reside on the finite 
components is zero.
Using a generating function formulation 
derived earlier to calculate the size distribution of the finite tree components
\cite{Newman2007,Kryven2017},
Wu et al. obtained the bridgeness distribution in configuration model
networks with Poisson, exponential
and power-law degree distributions
\cite{Wu2018}.
It was found that the mean bridgeness diverges
at $c \rightarrow c_0^{+}$ and
and monotonically decreases as the mean degree is increased.

Another useful measure of the importance of an edge $e$ in a network
is given by its range $\rho$, which is the distance between its end-nodes
$i$ and $i'$ in the reduced network from which $e$ is removed
\cite{Granovetter1973,Granovetter1983}.
In the special case in which $e$ is a bredge, its range is $\rho=\infty$,
because upon deletion of $e$ its end-nodes land on different
network components. For edges that are not bredges the range 
$\rho \ge 2$ is finite. It is equal to the shortest path length between
$i$ and $i'$ in the reduced network. It also satisfies $\rho = \ell-1$ where
$\ell$ is the length of the shortest cycle that includes the edge $e$
in the original network. Edges whose range $\rho$ is large are considered
important because upon their removal the shortest alternate path 
between $i$ and $i'$ is large. In practical applications, large $\rho$ 
implies long and potentially costly delays in communication and 
transportation in case that the edge $e$ fails.

\begin{figure}
\includegraphics[width=7.5cm]{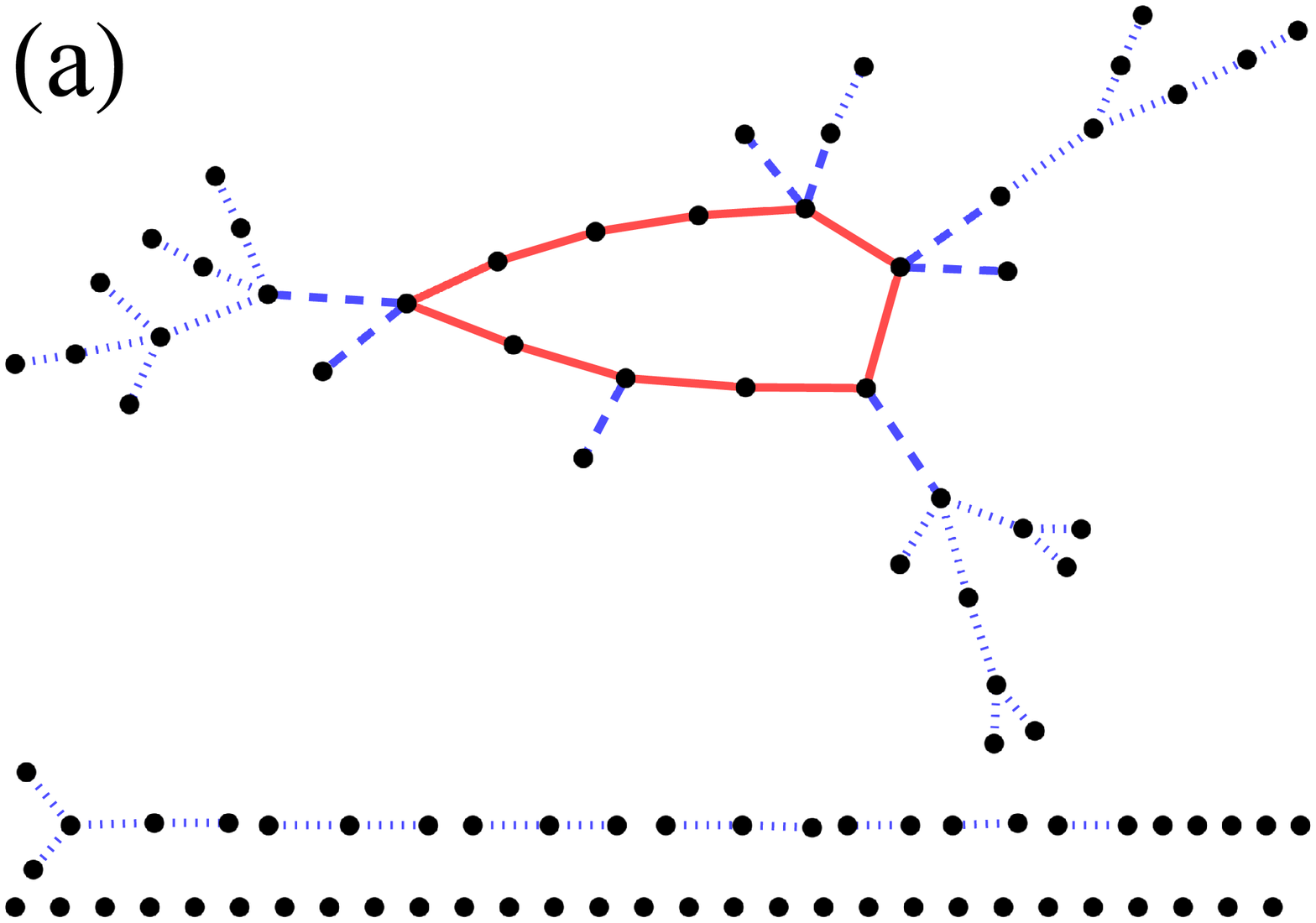}
\\  
\vspace{0.8in}
\includegraphics[width=7.5cm]{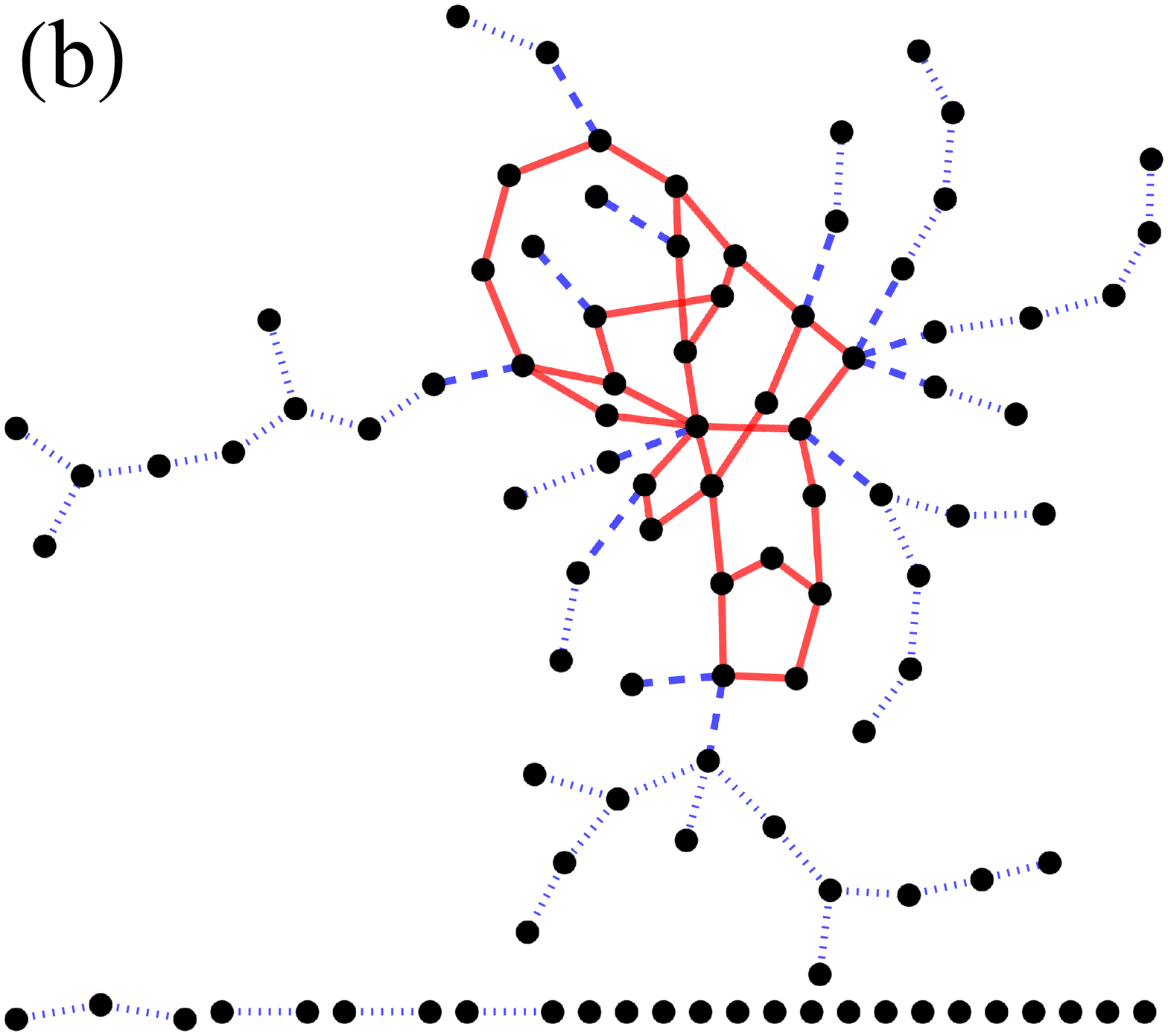}
\caption{
(Color online)
ER networks of $N=100$ nodes with mean degree $c=1.1$ (a) and $c=1.7$ (b),
which exhibit a coexistence between a giant component 
and finite tree components.
The non-bredge edges (solid lines) connect pairs of nodes that reside on
the 2-core of the giant component. 
In the more dilute case (a) the 2-core consists of a single cycle,
while in the denser case (b) it exhibits a complex web of cycles.
The root bredges (dashed lines) connect the tree branches on the giant
component to the 2-core. All the other bredges (dotted lines) connect pairs of nodes on that reside on the tree branches  
of the giant component and pairs of nodes on the finite tree components.
}
\label{fig:14}
\end{figure}

In Fig. \ref{fig:14} we present ER networks of $N=100$ nodes with
mean degrees $c=1.1$ [Fig. \ref{fig:14}(a)] and $c=1.7$ [Fig. \ref{fig:14}(b)].
In both networks the giant component coexists with many finite components.
The non-bredge edges (solid lines) connect pairs of nodes that reside
on the 2-core of the giant component
\cite{Newman2008,Dufresne2013}. 
The giant component is decorated by tree branches,
on which all the edges are bredges.
The bredge that connects each tree branch to the 2-core of the giant
component is called root bredge (dashed line).
The end-node of the root bredge that resides on the 2-core
is called root end-node.
All the other bredges (dotted lines) connect pairs of nodes
that reside on the tree branches, 
which are not on the 2-core.
The average size of the tree branches that decorate the giant component
is given by
\cite{Wu2018}
\begin{equation}
N_{\rm T} = \frac{1}{1 - G'_1(1-\tilde g)},
\end{equation}

\noindent
which is the sum of a geometric series whose ratio $G'_1(1-\tilde g)$
is the excess degree of the end-nodes of the finite component
side of the bredges, whose degree distribution is given by 
Eq. (\ref{eq:tPKFk'BGC}).
Thus, the fraction of root end-nodes among the end-nodes on
the GC side of bredges on the giant component is $1/N_{\rm T}$.
The degree distribution of the root end-nodes,
which reside on the 2-core of the giant component, 
is given by
\begin{eqnarray}
\widetilde P(K_{ \scriptsize{\mbox{2-CORE}} }=k|{\rm B},{\rm GC}) & \nonumber\\
& \hspace{-25mm}=
\frac{
\left[ 1 - (1-\tilde g)^{ k-1}  \right]
- (k-1) \tilde g(1-\tilde g)^{k-2}
}
{\tilde g [1 - G'_1(1-\tilde g)] }
\widetilde P(k).
\label{eq:tPKGkBGC2}
\end{eqnarray}

\noindent
The degree distribution of the 
end-nodes 
on the GC sides
of all other bredges,
which reside on the 1-core
of the giant component
is given by
\begin{equation}
\widetilde P(K_{ {\rm GC} \cap \overline{ \scriptsize{\mbox{2-CORE} }} }=k|{\rm B},{\rm GC}) =
\frac{ (k-1) (1-\tilde g)^{k-2} }{G'_1(1-\tilde g)}
\widetilde P(k).
\label{eq:tPKGk'BGC1}
\end{equation}

\noindent
The overall distribution of the degrees $K_{\rm GC}$,
given by Eq. (\ref{eq:tPKGkBGC}), 
is recovered by
\begin{eqnarray}
\widetilde P(K_{\rm GC}&=&k|{\rm B},{\rm GC}) =
\frac{1}{N_{\rm T}}  
\widetilde P(K_{  \scriptsize{\mbox{2-CORE}} }=k|{\rm B},{\rm GC}) \nonumber\\
& & \hspace{-10mm}+
\left( 1 - \frac{1}{N_{\rm T}} \right)
\widetilde P(K_{ {\rm GC} \cap \overline{ \scriptsize{\mbox{2-CORE} }}  }=k |{\rm B},{\rm GC}).
\label{eq:tPKGkBGC3}
\end{eqnarray}

\noindent
The distinction between root bredges and all the other bredges on the
giant component may be useful for optimized dismantling algorithms and
targeted attacks. This is due to the fact that the deletion of a root bredge
disconnects the whole tree branch that is held by this bredge.
In contrast, random deletion of bredges may require a large number of
deletion steps in order to chop each tree branch from the 2-core of
the giant component.

\section{Summary}

We presented analytical results for the statistical properties of 
edges and bredges in configuration model networks.
To quantify the abundance of bredges,
we calculated the probability
$\widehat P(e \in {\rm B})$
that a random edge $e$ in a configuration model network 
with a given degree distribution $P(k)$
is a bredge.
We also obtained the conditional probability 
$\widehat P(e \in {\rm B} | k,k')$
that a random
edge whose end-nodes are of degrees $k$ and $k'$ is a bredge.
Using Bayes' theorem, we obtained the joint degree distribution  
$\widehat P(k,k' | {\rm B})$
of the end-nodes of randomly sampled bredges.
We also studied the distinct properties of bredges on the giant component
and on the finite components.
On the finite components all the edges are bredges,
namely $\widehat P(e \in {\rm B}|{\rm GC})=1$, and there
are no degree-degree correlations.
We calculated the probability $\widehat P(e \in {\rm B}|{\rm GC})$
that a random edge on the giant component is a bredge.
We also obtained the joint degree distribution
$\widehat P(k,k'|{\rm B},{\rm GC})$
of the end-nodes of bredges
and the joint degree distribution 
$\widehat P(k,k'|{\rm NB},{\rm GC})$
of the end-nodes of non-bredge (NB) edges
on the giant component.
Surprisingly, it was found that the degrees $k$ and $k'$ of the end-nodes of bredges are correlated,
while the degrees of the end-nodes of non-bredge edges are uncorrelated.
This implies that all the degree-degree correlations on the giant component
are concentrated on the bredges.
We calculated the
covariance
$\Gamma({\rm B},{\rm GC})$ 
and found that it is negative, which means that bredges 
on the giant component tend
to connect high degree nodes to low degree nodes and vice versa.
We applied this analysis to ensembles of configuration model networks
with degree distributions that follow 
a Poisson distribution (Erd{\H o}s-R\'enyi networks), 
an exponential distribution 
and a power-law distribution 
(scale-free networks).
The implications of these results were discussed in the context of common
attack scenarios and network dismantling processes.

This work was supported by the Israel Science Foundation grant no. 
1682/18.

\clearpage
\newpage


\clearpage
\newpage





\begin{thebibliography}{10}


\bibitem{Havlin2010}
S. Havlin and R. Cohen,
{\it Complex Networks: Structure, Robustness and Function}
(Cambridge University Press, New York, 2010).

\bibitem{Newman2010}
M.E.J. Newman, 
{\it Networks: an Introduction, 1st Edition} 
(Oxford University Press, Oxford, 2010).

\bibitem{Estrada2011}
E. Estrada, 
{\it The Structure of Complex Networks: Theory and Applications} 
(Oxford University Press, Oxford, 2011).

\bibitem{Barrat2012}
A. Barrat, M. Barth\'elemy and A. Vespignani,
{\it Dynamical Processes on Complex Networks}
(Cambridge University Press, Cambridge, 2012).


\bibitem{Latora2017}
V. Latora, V. Nicosia, G. Russo,
{\it Complex Networks: Principles, Methods and Applications},
(Cambridge University Press, Cambridge, 2017).


\bibitem{Bollobas2001}
B. Bollob\'as, 
{\it Random Graphs, 2nd Edition}
(Cambridge University Press, Cambridge, 2001).



\bibitem{Albert2000}
R. Albert, H. Jeong and A.-L. Barab\'{a}si,
Error and attack tolerance of complex networks,
{\it Nature} {\bf 406}, 378 (2000).

\bibitem{Cohen2000}
R. Cohen, K. Erez, D. ben-Avraham and S. Havlin,
Resilience of the internet to random breakdowns,
{\it Phys. Rev. Lett.} {\bf 85}, 4626 (2000).

\bibitem{Cohen2001}
R. Cohen, K. Erez, D. ben-Avraham and S. Havlin,
Breakdown of the internet under intentional attack,
{\it Phys. Rev. Lett.} {\bf 86}, 3682 (2001).

\bibitem{Schneider2011}
C.M. Schneider, A.A. Moreira, J.S. Andrade, S. Havlin and H.J. Herrmann,
Mitigation of malicious attacks on networks,
{\it Proc. Natl. Acad. Sci. USA} {\bf 108}, 3838 (2011).


\bibitem{Hopcroft1973}
J. Hopcroft	and R. Tarjan,
Efficient algorithms for graph manipulation,
{\it Communications of the ACM}
{\bf 16}, 372 (1973).

\bibitem{Gibbons1985}
A. Gibbons,
{\it Algorithmic Graph Theory}
(Cambridge University Press, Cambridge, 1985).




\bibitem{Chaudhuri1998}
P. Chaudhuri,
An optimal distributed algorithm for finding afticulation points in a network,
{\it Computer Communications} {\bf 21}, 1707 (1998).






\bibitem{Tian2017}
L. Tian, A. Bashan, D.-N. Shi and Y.-Y. Liu,
Articulation points in complex networks,
{\it Nature Communications} {\bf 8}, 14223 (2017).

\bibitem{Italiano2012}
G.F. Italiano, L. Laura and F. Santaroni,
Finding strong bridges and strong articulation points in linear time, 
{\it Theoretical Computer Science} {\bf 447}, 74 (2012).


\bibitem{Tarjan1974}
R.E. Tarjan, 
A note on finding the bridges of a graph, 
{\it Information Processing Letters} {\bf 2}, 160 (1974).


\bibitem{Bollobas1998}
B. Bollobas,
{\it Modern Graph Theory}
(Springer, New York, 1998).


\bibitem{Bredge1685}
The word 'bredge' is an obsolete form of the word 'bridge'
in ancient English;
See e.g. 
English Dictionary, explaining the difficult terms,
by E. Coles, School-Master and Teacher,
printed for Peter Parker at the Leg and Star 
over against the Royal-Exchange in Cornbil (1685).

\bibitem{Braunstein2016}
A. Braunstein, L. Dall'Asta, G. Semerjian and L. Zdeborov\'a,
Network dismantling,
{\it Proc. Natl. Acad. Sci. USA} {\bf 113}, 12368 (2016). 


\bibitem{Zdeborova2016}
L. Zdeborov\'a, P. Zhang and H.-J. Zhou,
Fast and simple decycling and dismantling of networks,
{\it Scientific Reports} {\bf 6}, 37954 (2016).


\bibitem{Wandelt2018}
S. Wandelt, X. Sun, D. Feng, M. Zanin and S. Havlin,
A comparative analysis of approaches to network-dismantling,
{\it Scientific Reports} {\bf 8}, 13513 (2018).

\bibitem{Ren2019}
X.-L. Ren, N. Gleinig, D. Helbing and N. Antulov-Fantulina,
Generalized network dismantling,
{\it Proc. Natl. Acad. Sci. USA} {\bf 116}, 6554 (2019).

\bibitem{Tishby2018a}
I. Tishby, O. Biham, R. K\"uhn and E. Katzav,
Statistical analysis of articulation points in configuration model networks,
{\it Phys. Rev. E} {\bf 98}, 062301 (2018).



\bibitem{Molloy1995}
M. Molloy and B. Reed,
A critical point for random graphs with a given degree sequence,
{\it Random Struct. Algorithms} {\bf 6}, 161 (1995).

\bibitem{Molloy1998}
M. Molloy and B. Reed,
The size of the giant component of a random graph 
with a given degree sequence,
{\it Combinatorics, Probability and Computing}
{\bf 7}, 295 (1998).


\bibitem{Newman2001}
M.E.J. Newman, S.H. Strogatz and D.J. Watts,
Random graphs with arbitrary degree distributions and their applications,
{\it Phys. Rev. E} {\bf 64}, 026118 (2001).




\bibitem{Erdos1960b}
P. Erd{\H o}s and T. Gallai, 
Gr\'afok el{\H o}\'irt foksz\'am\'u pontokkal,
{\it Matematikai Lapok} {\bf 11}, 264 (1960).

\bibitem{Choudum1986}
S.A. Choudum, 
A simple proof of the Erd{\H o}s-Gallai theorem on graph sequences, 
{\it Bulletin of the Australian Mathematical Society} {\bf 33},  67 (1986).



\bibitem{Bonneau2017}
H. Bonneau, A. Hassid, O. Biham, R. K\"uhn and E. Katzav, 
Distribution of shortest cycle lengths in random networks, 
{\it Phys. Rev. E} {\bf 96}, 062307 (2017).




\bibitem{Mezard1985}
M. M\'ezard, G. Parisi and M.A. Virasoro, 
Random free energies in spin glasses,
{\it J. Physique Lett.} {\bf 46}, L217 (1985).

\bibitem{Mezard2003}
M. M\'ezard and G. Parisi, 
The cavity method at zero temperature,
{\it J. Stat. Phys.} {\bf 111}, 1 (2003).

\bibitem{Mezard2009}
M. M\'ezard and A. Montanari,
{\it Information, Physics and Computation}
(Oxford University Press, Oxford, 2009).

\bibitem{Ferraro2015}
G. Del Ferraro, C. Wang, D. Mart\'i and M. M\'ezard, 
Cavity method - message passing from a physics perspective, 
{\it Statistical Physics, Optimization, 
Inference and Message-Passing Algorithms},
Lecture Notes of the Les Houches School of Physics, 
Eds. F. Krzakala, F. Ricci-Tersenghi, L. Zdeborova, R. Zecchina, 
E.W. Tramel, and L.F. Cugliandolo 
(Oxford University Press, Oxford, 2015).




\bibitem{Lowy2004}
A. Lowy and P. Wood,
{\it The power of the $2 \times 2$ matrix:
using $2 \times 2$ thinking to solve business problems and 
make better decisions}
(Jossey Bass, San Francisco, 2004)


\bibitem{Tishby2018b}
I. Tishby, O. Biham, E. Katzav and R. K\"uhn,
Revealing the microstructure of the giant component in random graph ensembles,
{\it Phys. Rev. E} {\bf 97}, 042318 (2018).



\bibitem{Tishby2019}
I. Tishby, O. Biham, E. Katzav and R. K\"uhn,
Generating random networks that consist of a single connected component with a given degree distribution,
{\it Phys. Rev. E} {\bf 99}, 042308 (2019).

\bibitem{Seidman1983}
S.B. Seidman,
Network structure and minimum degree,
{\it Social Networks} {\bf 5}, 269 (1983).


\bibitem{Yuan2016}
X. Yuan, Y. Dai, H.E. Stanley and S. Havlin,
k-core percolation on complex networks: Comparing random, localized, and targeted attacks,
{\it Phys. Rev. E} {\bf 93}, 062302 (2016).


\bibitem{Dorogovtsev2006a}
S.N. Dorogovtsev, A.V. Goltsev and J.F.F. Mendes, 
k-Core Organization of Complex Networks,
{\it Phys. Rev. Lett.} {\bf 96}, 040601 (2006).



\bibitem{Dorogovtsev2006b}
S.N. Dorogovtsev, A.V. Goltsev and J.F.F. Mendes,
k-core architecture and k-core percolation on complex networks,
{\it Physica D} {\bf 224}, 7 (2006).




\bibitem{Diekmann2000}
O. Diekmann and J.A.P. Heesterbekk,
{\it Mathematical Epidemiology of Infectious Diseases: Model Building, Analysis and Interpretation}
(John Wiley \& Sons, Chichester, 2000).


\bibitem{Duderstadt1976}
J. Duderstadt and L. Hamilton, 
{\it Nuclear Reactor Analysis}
(John Wiley \& Sons, USA, 1976)


\bibitem{Katzav2018}
E. Katzav, O. Biham and A. Hartmann,
Metric properties of subcritical Erd{\H o}s-R\'enyi networks,
{\it Phys. Rev. E} {\bf 98}, 012301 (2018).

\bibitem{Newman2002b}
M.E.J. Newman,
Assortative mixing in networks,
{\it Phys. Rev. Lett.} {\bf 89}, 208701 (2002).


\bibitem{Newman2003a}
M.E.J. Newman,
Mixing patterns in networks,
{\it Phys. Rev. E} {\bf 67}, 026126 (2003).


\bibitem{Newman2003b}
M.E.J. Newman and J. Park,
Why social networks are different from other types of networks,
{\it Phys. Rev. E} {\bf 68}, 036122 (2003).

\bibitem{Johnson2010}
S. Johnson, J.J. Torres, J. Marro and M.A. Munoz,
Entropic Origin of Disassortativity in Complex Networks,
{\it Phys. Rev. Lett.} {\bf 104},  108702  (2010).



\bibitem{Mizutaka2018}
S. Mizutaka and T. Hasegawa,
Disassortativity of percolating clusters in random networks,
{\it Phys. Rev. E} {\bf 98}, 062314 (2018).



\bibitem{Wu2018}
A.-K. Wu, L. Tian and Y.-Y. Liu,
Bridges in complex networks,
{\it Phys. Rev. E} {\bf 97}, 012307 (2018).




\bibitem{Erdos1959}
P. Erd{\H o}s and R\'{e}nyi, 
On random graphs I,
{\it Publicationes Mathematicae} {\bf 6}, 290 (1959).

\bibitem{Erdos1960}
P. Erd{\H o}s and R\'{e}nyi, 
On the evolution of random graphs
{\it Publ. Math. Inst. Hung. Acad. Sci.} 
{\bf 5}, 17 (1960).

\bibitem{Erdos1961}
P. Erd{\H o}s and R\'{e}nyi,
On the evolution of random graphs II
{\it Bull. Inst. Int. Stat.} 
{\bf 38}, 343 (1961).


\bibitem{Olver2010}
F.W.J. Olver, D.M. Lozier, R.F. Boisvert and C.W. Clark,
{\it NIST Handbook of Mathematical Functions} 
(Cambridge University Press, Cambridge, 2010).



\bibitem{Gradshteyn2000}
I.S. Gradshteyn and I.M. Ryzhik, 
{\it Tables of Integrals, Series, and Products}, 
6th edition 
(Academic Press, San Diego, 2000).



\bibitem{Newman2008}
M.E.J. Newman and G. Ghoshal,
Bicomponents and the robustness of networks to failure,
{\it Phys. Rev. Lett.} {\bf 100}, 138701 (2008).

\bibitem{Dufresne2013}
L. H\'ebert-Dufresne, A. Allard, J.-G. Young, and L.J. Dub\'e,
Percolation on random networks with arbitrary k-core structure,
{\it Phys. Rev. E} {\bf 88}, 062820 (2013).



\bibitem{Chung2003}
F. Chung and L. Lu, 
The average distance in a random graph with given expected degrees,
{\it Internet Mathematics} {\bf 1}, 91 (2003).

\bibitem{Fronczak2004}
A. Fronczak, P. Fronczak, and J.A. Holyst, 
Average path length in random networks,
{\it Phys. Rev. E} {\bf 70}, 056110 (2004).



\bibitem{Durrett2007}
R. Durrett,
{\it Random Graph Dynamics}
(Cambridge University Press, Cambridge, 2007)


\bibitem{Hartmann2017}
A.K. Hartmann, M. M\'ezard,
{\it Phys. Rev. E} {\bf 97}, 032128 (2017)


\bibitem{Dorogovtsev2003}
S.N. Dorogovtsev, J.F.F. Mendes and A.N. Samukhin, 
Metric structure of random networks,
{\it Nuclear Physics B} {\bf 653}, 307 (2003).







\bibitem{Blondel2007}
V.D. Blondel, J.-L. Guillaume, J.M. Hendrickx and R.M. Jungers, 
Distance distribution in random graphs and application to network exploration,
{\it Phys. Rev. E} {\bf 76}, 066101 (2007).


\bibitem{Hofstad2008}
R. van der Hofstad and G. Hooghiemstra, 
Universality for distances in power-law random graphs,
{\it J. Math. Phys.} {\bf 49}, 125209 (2008).

\bibitem{Esker2008}
H. van der Esker, R. van der Hofstad, G. Hooghiemstra, 
Universality for the Distance in Finite Variance Random Graphs,
{\it J. Stat. Phys.} {\bf 133}, 169 (2008).

\bibitem{Shao2008}
J. Shao, S.V. Buldyrev, R. Cohen, M. Kitsak, S. Havlin and H.E. Stanley,
{\it EPL} {\bf 84}, 48004 (2008).

\bibitem{Shao2009}
J. Shao, S.V. Buldyrev, L.A. Braunstein, S. Havlin and H.E. Stanley,
Structure of shells in complex networks
{\it Phys. Rev. E} {\bf 80}, 036105 (2009).

\bibitem{Katzav2015}
E. Katzav, M. Nitzan, D. ben-Avraham, P.L. Krapivsky, 
R. K\"uhn, N. Ross and O. Biham,
Analytical results for the distribution of shortest path lengths in random networks,
{\it EPL} {\bf 111}, 26006 (2015).



\bibitem{Nitzan2016}
M. Nitzan, E. Katzav, R. K\"uhn and O. Biham,
Distance distribution in configuration-model networks,
{\it Phys. Rev. E} {\bf 93}, 062309 (2016).





\bibitem{Melnik2016}
S. Melnik and J.P. Gleeson, 
Simple and accurate analytical calculation of shortest path lengths,
arXiv:1604.05521. 



\bibitem{Asher2020}
E.E. Asher, H. Sanhedrai, N.K. Panduranga, R. Cohen and S. Havlin,
Distance distribution in extreme modular networks,
{\it Phys. Rev. E} {\bf 101}, 022313 (2020)


\bibitem{Steinbock2017}
C. Steinbock, O. Biham and E. Katzav,
The distribution of shortest path lengths in a class of node duplication network models,
{\it Phys. Rev. E} {\bf 96}, 032301 (2017).

\bibitem{Steinbock2019a}
C. Steinbock, O. Biham and E. Katzav,
Analytical results for the distribution of shortest path lengths in directed random networks that grow by node duplication,
{\it Europ. Phys. J. B} {\bf 92}, 130 (2019).

\bibitem{Steinbock2019b}
C. Steinbock, O. Biham and E. Katzav,
Analytical results for the in-degree and out-degree distributions of directed random networks that grow by node duplication,
{\it J. Stat. Mech.}  083403 (2019).


\bibitem{Freeman1977}
L.C. Freeman,
A set of measures of centrality based on betweenness,
{\it Sociometry} {\bf 40}, 35 (1977).
 
\bibitem{Goh2003}
K.I. Goh, E. Oh, B. Kahng and D. Kim,
Betweenness centrality correlation in social networks
{\it Phys. Rev. E} {\bf 67}, 017101 (2003).



\bibitem{Brandes2001}
U. Brandes,
A faster algorithm for betweenness centrality,
{\it Journal of Mathematical Sociology} {\bf 25}, 163 (2001).

\bibitem{Newman2007}
M.E.J. Newman,
Component sizes in networks with arbitrary degree distributions
{\it Phys. Rev. E} {\bf 76}, 045101 (2007).

\bibitem{Kryven2017}
I. Kryven,
General expression for the component size distribution in infinite configuration networks,
{\it Phys. Rev. E} {\bf 95}, 052303 (2017).


\bibitem{Granovetter1973}
M. Granovetter,
The strength of weak ties,
{\it American Journal of Sociology} {\bf 78}, 1360 (1973).

\bibitem{Granovetter1983}
M. Granovetter,
The strength of weak ties: a network theory revisited,
{\it Sociological Theory} {\bf 1}, 201 (1983).

\end{thebibliography}
\end{document}